\documentclass{natureS}

\usepackage{bm}
\usepackage{graphicx}
\usepackage{amssymb,amsmath}
\usepackage{upgreek}
\usepackage{float}

\usepackage[resetlabels]{multibib}

\usepackage{booktabs,multirow}
\usepackage[table]{xcolor}
\newcites{S}{Supplementary References}

\newcommand{\beginsupplement}{%
        \setcounter{table}{0}
        \renewcommand{\tablename}{Supplementary Table}
        \setcounter{figure}{0}
        \renewcommand{\figurename}{Supplementary Figure}
        \setcounter{section}{0}
        \renewcommand{\thesection}{Supplementary Note \arabic{section}}%
        \setcounter{subsection}{0}
        \renewcommand{\thesubsection}{S\arabic{section}.\arabic{subsection}}%
        \setcounter{equation}{0}
     }
\makeatletter
\let\saved@includegraphics\includegraphics
\AtBeginDocument{\let\includegraphics\saved@includegraphics}
\renewenvironment*{figure}{\@float{figure}}{\end@float}
\makeatother     
\definecolor{applegreen}{rgb}{0.29, 0.6, 0.0}

\title{\begin{center} Oxygen vacancy-driven orbital multichannel Kondo effect in Dirac nodal line metals IrO$_2$ and RuO$_2$
 \end{center}}

\author{Sheng-Shiuan Yeh,$^{1,2,3}$ Ta-Kang Su,$^1$ An-Shao Lien,$^1$ Farzaneh Zamani,$^4$ Johann Kroha,$^4$\\ Chao-Ching Liao,$^1$ Stefan Kirchner,$^{5,6,\ast}$ and Juhn-Jong Lin$^{1,2,7,\ast}$}

\begin{document}

\maketitle

\begin{affiliations}
 \item NCTU-RIKEN Joint Research Laboratory, Institute of Physics, National Chiao Tung \mbox{University}, Hsinchu 30010, Taiwan
 \item Center for Emergent Functional Matter Science, National Chiao Tung University, Hsinchu 30010, Taiwan
 \item International College of Semiconductor Technology, National Chiao Tung University, Hsinchu 30010, Taiwan 
 \item Physikalisches Institut and Bethe Center for Theoretical Physics, Universit\"{a}t Bonn, Nussallee 12, D-53115 Bonn, Germany
 \item Zhejiang Institute of Modern Physics and Department of Physics, Zhejiang University, Hangzhou, 310027, China
 \item Zhejiang Province Key Laboratory of Quantum Technology and Device, Zhejiang University, Hangzhou 310027, China
\item Department of Electrophysics, National Chiao Tung University, Hsinchu 30010, Taiwan\\
$^\ast$E-mail: 
stefan.kirchner@correlated-matter.com, jjlin@mail.nctu.edu.tw
\end{affiliations}

\newpage 

\noindent{\bf ABSTRACT}\\
{\bf Strong electron correlations  have long been recognized as driving the emergence of novel phases of matter. A well recognized example is high-temperature superconductivity which cannot be understood in terms of the standard weak-coupling theory. The exotic properties that accompany the formation of the two-channel Kondo (2CK) effect  including the emergence of  an unconventional metallic state  in the low-energy limit also originate from strong electron interactions. Despite its paradigmatic role for the formation of non-standard metal behavior, the stringent conditions required for its emergence have made the observation of the nonmagnetic, orbital 2CK effect in real quantum materials difficult, if not impossible. We report the observation of orbital one- and two-channel Kondo physics in the symmetry-enforced Dirac nodal line (DNL) metals IrO$_2$ and RuO$_2$ nanowires and show that the symmetries that enforce the existence of DNLs also promote the formation of nonmagnetic Kondo correlations. Rutile oxide nanostructures thus form a versatile  quantum matter platform to engineer and explore intrinsic, interacting topological states of matter.
}

\begin{figure}
	\centering
	\includegraphics[width=1.0\linewidth]{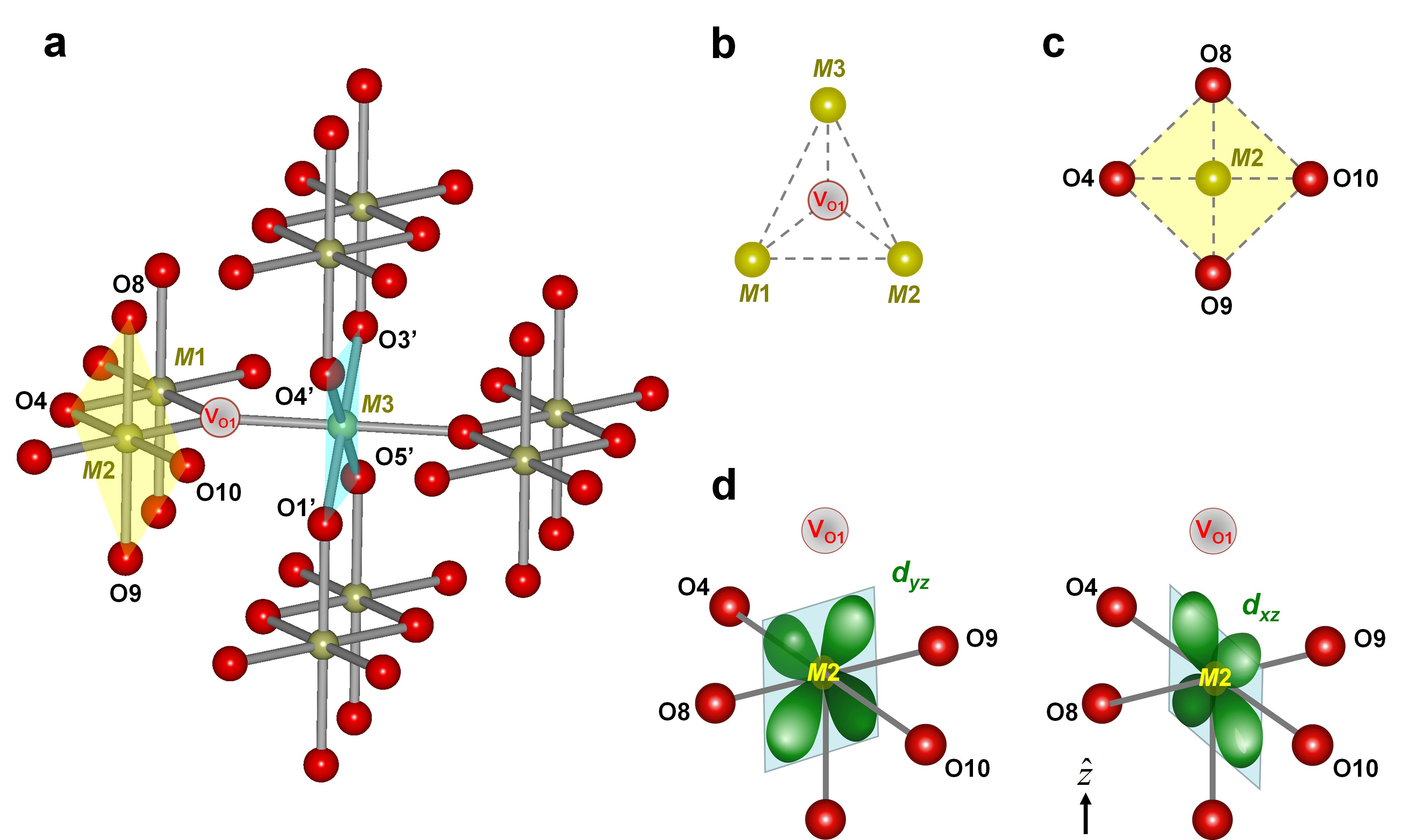}
	\caption{\textbf{Atomic arrangement around an oxygen vacancy in $M$O$_2$ rutile structure.}
		\textbf{a} Schematics for $M$O$_2$ in the rutile structure.  The olive and red spheres represent transition-metal ions $M^{4+}$ and oxygen ions O$^{2-}$, respectively. V$_{\rm O1}$ represents an oxygen vacancy. \textbf{b} The metal ions $M$1, $M$2 and $M$3 surrounding V$_{\rm O1}$ form an isosceles triangle. \textbf{c} The four oxygen ions surrounding $M$2, labeled O4, O9, O10 and O8, form an almost perfect planar square (while O1$^\prime$, O5$^\prime$, O3$^\prime$, and O4$^\prime$ only form a rectangle, cf. \ref{sec:S-defectStructure} for details). \textbf{d} The $d_{xz}$ and $d_{yz}$ orbitals at $M$2 next to V$_{\rm O1}$, with $\hat{z}$ perpendicular to the O4, O8, O9 and O10 plane, remain  essentially degenerate as a result of mirror and $C_4$ rotation  symmetry around $M$2. (Due to the non-symmorphic rutile structure, $\hat{z}\nparallel \hat{z}^{\prime}$,  where $\hat{z}^{\prime}$ is parallel to the $C_4$ axis at $M$1.)
	}
	\label{fig_1}
\end{figure}

\begin{figure}
	\centering
	\includegraphics[width=0.6\linewidth]{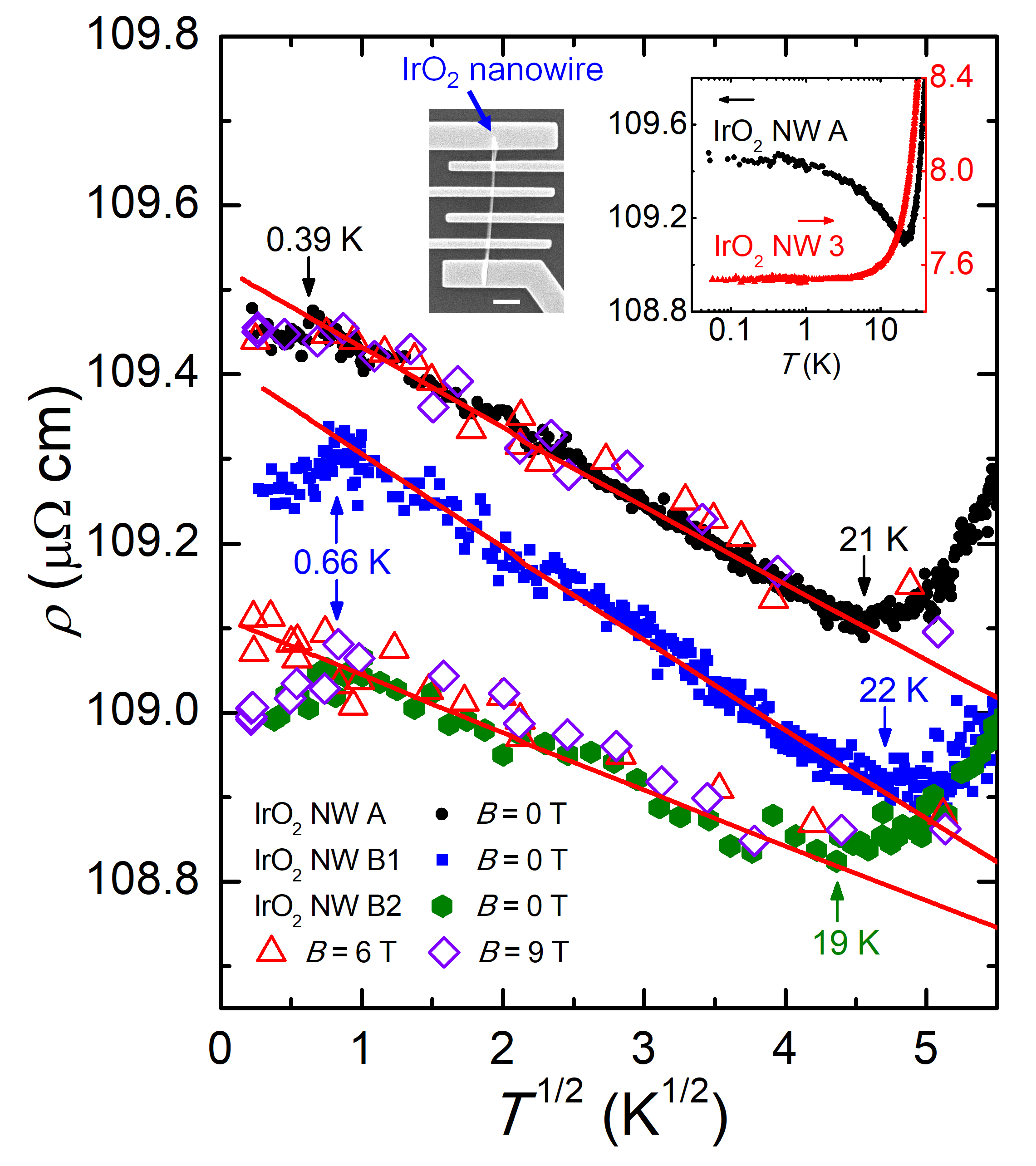}
	\caption{\textbf{Orbital 2CK resistivity of IrO$_2$ NWs.}
		$\rho$ versus $\sqrt{T}$ for IrO$_2$ NWs A, B1 and B2 in magnetic fields $B$ = 0, 6 and 9 T, as indicated. For clarity, the data of NWs B1 and B2 are shifted by 34.7 and 33.6 $\upmu \Omega$ cm, respectively. A $\rho\propto \sqrt{T}$ law, which is $B$ independent, is observed between $\sim$\,0.5 and $\sim$\,20 K in all three NWs. The straight solid lines are linear fits to the 2CK resistivities calculated by the dynamical large-$N$ method (see text). Top left inset: A scanning electron microscopy image of NW A. The scale bar is 1 $\upmu$m. Top right inset: Low-$T$ $\rho(T)$ curves of NW A and a reference, oxygenated NW 3 (diameter $d=330$ nm, $\rho$(300\,K) = 124 $\upmu\Omega$ cm). 
	}
	\label{fig_2}
\end{figure}

\begin{introduction}
Unconventional metallic states and the breakdown of the Landau Fermi liquid paradigm is a central topic in contemporary condensed matter science. A connection with high-temperature superconductivity is experimentally well established but the conditions under which these enigmatic metals form has remained perplexing\cite{Keimer.15}. One of the simplest routes to singular Fermi liquid behavior, at least conceptually, is through two-channel Kondo (2CK) physics\cite{Nozieres.80,Vladar.83a,Cox.98}. Despite this long-standing interest, 2CK physics has thus far only been demonstrated to arise in carefully designed semiconductor nanodevices in narrow energy and temperature ($T$) ranges\cite{Potok.07,Keller.15,Iftikhar.15,Iftikhar.18}, while claims of its observation in real quantum materials are contentious (see DISCUSSION section for details).
More recently, the interest in Dirac and Weyl fermions within  a condensed matter framework has led to the exploration of the effects of strong spin-orbit coupling (SOC) and of topological states which are rooted in a combination of time-reversal, particle-hole, and space-group symmetries\cite{Burkov.16,Yang.18}. While there has been considerable progress in understanding weakly correlated topological metals, only a few materials have been identified as realizing topological phases driven by strong electron correlations, which includes the Weyl-Kondo semimetals\cite{Dzsaber.17}.
This raises the question if the 2CK counterpart of such a Weyl-Kondo semimetal, featuring an entangled ground state of the low-energy excitations of the 2CK effect with band-structure enforced Dirac or Weyl excitations,  could at least in principle be stabilized.
Exploring such a possibility, however, hinges on whether the 2CK effect can be stabilized at all in native quantum matter. 

In this work we establish that oxygen vacancies (V$_{\rm{\small O}}$'s) in the Dirac nodal line (DNL) materials IrO$_2$ and RuO$_2$ drive an orbital Kondo effect. V$_{\rm{\small O}}$'s are prevalent in transition metal oxides, including, e.g., TiO$_2$ and SrTiO$_3$, and their properties and ramifications have become  central research topics as they can lead to an intricate entanglement of spin, orbital, and charge degrees of freedom\cite{Ourmazd.87,Lin.13,Lin.15,Lechermann.17}.  The active degree of freedom in the orbital Kondo effect is not a local spin moment but a `pseudospin' formed by orbital degrees of freedom\cite{Cox.98}.
In IrO$_2$ and RuO$_2$, the  orbital Kondo effect is symmetry stabilized by the space-group symmetries of the rutile structure (Fig. \ref{fig_1}). Both materials have been characterized as topological metals which feature symmetry-protected DNLs in their Brillouin zones\cite{Jovic.18,Nelson.19}.
This provides a link between the formation of the orbital Kondo effect and the presence of DNLs. In IrO$_2$ a nonmagnetic 2CK ground state ensues, while in RuO$_2$ the absence of time-reversal symmetry results in an orbital one-channel Kondo (1CK) effect.

The rutile structure type possesses mirror reflection, inversion, and a fourfold rotation ($C_4$) symmetry which enforce the presence of DNLs in the band structure of rutile oxides\cite{Yang.18}. Some of these DNLs are protected from gapping out due to large SOC by the non-symmorphic symmetry of the rutile structure\cite{Zhao.16,Sun.17}. For IrO$_2$ and RuO$_2$ this has been recently confirmed by angle-resolved photoemission spectroscopy (ARPES) and band structure studies\cite{Sun.17,Jovic.18,Nelson.19}. In the vicinity of V$_{\rm{\small O}}$'s, this set of symmetries promotes the formation of the  orbital 1CK and 2CK effect. The emergent Majorana zero mode that accompanies the formation of the 2CK effect is reflected in a singular excitation spectrum above the ground state which generates a $\sqrt{T}$-dependence of the resistivity $\rho(T)$ below a low-$T$ energy scale\cite{Coleman.95}, the Kondo temperature $T_{\rm{\small K}}$. This requires a well-balanced competition of two otherwise independent and degenerate screening channels and makes the 2CK effect extremely difficult to realize, especially in a natural quantum material\cite{Cox.98,Moustakas.97,Aleiner.02}. 
If one channel dominates over the other, the low-$T$ behavior will be that of conventional fermions. If the 2CK state arises out of  orbital Kondo scattering, magnetic-field ($B$) independence is expected for field strengths well above $T_{\rm{\small K}}$ as long as $g \mu_{\rm{\small B}} B\ll W$, where $g$ is the Land\'{e} factor, $\mu_{\rm{\small B}}$ is the Bohr magneton, and $W$ is the conduction electron half-bandwidth.
Our study is based on rutile ($M$O$_2$, $M=$ Ir, Ru) nanowires (NWs) which allow us to combine a high degree of sample characterization with an exceptional measurement sensitivity while probing material properties in the regime where the characteristic sample dimension is much larger than the elastic electron mean free path (cf. \ref{sec:S-EEI}). That is, we are concerned with weakly-disordered, diffusive metals which are three-dimensional (3D) with respect to the Boltzmann transport, whereas strong correlation effect causes a resistivity anomaly at low $T$. Table \ref{table_1} lists the relevant parameters for the NWs studied in this work.
\end{introduction}
\begin{results}

{\bf Oxygen vacancies in transition metal rutiles $M$O$_2$.}
In Fig.\ \ref{fig_1}{\bf a}, the vicinity of an V$_{\rm{\small O}}$, denoted V$_{\rm{\small O1}}$, is shown. The metal ions surrounding V$_{\rm{\small O1}}$, labeled  $M$1, $M$2 and $M$3, form an isosceles triangle (Fig.\ \ref{fig_1}{\bf b}). For the sites $M$1 and $M$2, an almost perfect $C_{4\nu}$ symmetry exists which implies a corresponding degeneracy associated with the two-dimensional irreducible representation of $C_{4\nu}$, see Fig.\ \ref{fig_1}{\bf c} and  \ref{sec:S-defectStructure}. In the pristine system, the metal ions are surrounded by oxygen octahedra anchored around the center and the corners of the tetragonal unit cell. The $\pi/2$ angle between adjacent octahedra leads to a fourfold screw axis symmetry. This non-symmorphic symmetry not only protects DNLs in IrO$_2$ against SOC-induced splitting\cite{Sun.17,Nelson.19}. It has also  been linked to the high electrical conductivity of IrO$_2$ (Ref. \citeonline{Yang.18}) and, as we find, is in line with the strong tendency to localize electrons  near V$_{\rm{\small O}}$'s required for the formation of orbital Kondo correlations. Moreover, the fourfold screw axis symmetry ensures that the $C_4$ rotation axes centered at the sites $M$1 and $M$2 near V$_{\rm{\small O1}}$ are not parallel ($\hat{z}^{\prime}\nparallel \hat{z}$, see Fig.\ \ref{fig_1}{\bf d}). This enhances the phase space for the orbital Kondo effect over orbital order linking sites $M$1 and $M$2 (see also \ref{sec:theory}).

{\bf Experimental signatures of orbital 2CK effect in IrO$_2$ NWs.} Now we turn to our experimental results which, to the best of our knowledge, demonstrate the most convincing realization of the long searched orbital 2CK effect in a solid. Figure \ref{fig_2} demonstrates the formation of an orbital 2CK effect in IrO$_2$ NWs. We find that as $T$ decreases from room temperature to $\sim$ a few  Kelvin, $\rho(T)$ decreases in all IrO$_2$ NWs, as expected for typical metallic behavior (cf. \ref{sec:S-rhoExtraction}). However, below $T$\,$\sim$\,20 K, $\rho(T)$ displays a $\sqrt{T}$ increase of the $\rho(T)$ upon lowering $T$ over almost two decades in $T$(!), until a deviation sets in at $\sim$\,0.5 K. We performed systematic thermal annealing studies to adjust the oxygen contents in the NWs, which indicate that the anomalous low-$T$ transport properties are driven by the presence of V$_{\rm{\small O}}$'s (Ref. \citeonline{Yeh.18} and  \ref{sec:S-annealing}). This is exemplified in Fig. \ref{fig_2}. The top left inset shows a scanning electron microscopy image of NW A. In the oxygenated NW 3 which should contain a negligible amount of V$_{\rm{\small O}}$'s, $\rho(T)$ decreases monotonically with decreasing $T$,
revealing a residual resistivity, $\rho_{\rm{\small B0}}$, below $\sim$\,4 K (top right inset). In contrast, in NWs A, B1 and B2 which contain large amounts of V$_{\rm{\small O}}$'s, $\rho(T)$ increases with decreasing $T$, manifesting a robust $\rho \propto \sqrt{T}$ law between $\sim$\,0.5 and $\sim$\,20 K. The slope of NW B2 is smaller than that of NW B1, which indicates a decrease in the number density of oxygen vacancies (n$_{\mbox{\tiny V$_{\mbox{\tiny O}}$}}$) due to prolonged aging (for about five months) in the atmosphere. The data explicitly demonstrate that the $\rho \propto \sqrt{T}$ behavior is independent of $B$ up to at least 9 T.
The observed behavior  is consistent with the 2CK effect as indicated by the straight solid lines which are linear fits to the 2CK effect calculated within the dynamical large-$N$ method (cf. \ref{sec:theory}), with n$_{\mbox{\tiny V$_{\mbox{\tiny O}}$}}$ as an adjustable parameter (see Table \ref{table_1} for the extracted values and  \ref{sec-SlargeN} and \ref{sec:S-numberdensity}  for the extraction method).

{\bf Ruling out the 3D electron-electron interaction (EEI) effect.} To complicate matters, the EEI effect in 3D weakly disordered metals generically leads to a $\sqrt{T}$ term in $\rho(T)$ at low $T$ (Refs. \citeonline{Altshuler.85} and \citeonline{Lee.85}). Unambiguously establishing that $\rho(T)\sim \sqrt{T}$ indeed originates from 2CK physics thus requires a proper analysis of the EEI effect of the charge carriers. For example, for the NW B1 with $\rho_{\rm{\small B0}}$ = 74 $\upmu \Omega$ cm and the electron diffusion constant $D \simeq$ 6.2 cm$^2$ s$^{-1}$, the 3D EEI effect would predict a largest possible resistance increase of $\triangle \rho/\rho \simeq 2.8 \times 10^{-4}$ as $T$ is cooled from 20 to 1 K. Experimentally, we have observed a much larger resistance increase of $5.1\times 10^{-3}$. Furthermore, the 3D EEI effect would predict similar values for the magnitude of the low-$T$ resistivity increase in NWs B1 and B2 to within $\approx$\,3\%, due to their $\rho_{\rm{\small B0}}$ values differing by $\approx$\,1\% (Table \ref{table_1}). This is definitely incompatible with our observation of a $\approx$\,50\% difference. In addition, we find a deviation from the $\sqrt{T}$ behavior at $\sim$\,0.5 K. If the $\sqrt{T}$ anomaly were caused by the EEI effect, no such deviation should occur (see  \ref{sec:S-EEI} for an in-depth analysis of the EEI effect and its 3D dimensionality in our $M$O$_2$ NWs).

{\bf V$_{\rm{\small O}}$-driven orbital Kondo scattering in $M$O$_2$.}
For IrO$_2$ the valency of the transition-metal ion $M$ is close to the nominal valence of +IV in $M$O$_2$ (Ref. \citeonline{Ping.15}).
Each V$_{\rm{\small O}}$  generates two {\itshape defect electrons} due to charge neutrality. To minimize Coulomb interaction, the {\itshape defect electrons} will tend to localize at different $M$ ions in the vicinity of the V$_{\rm{\small O}}$. In IrO$_2$ this results in a nonmagnetic 5$d^6$ ground state configuration of the Ir ions. For the electron localizing on ion $M$2 or $M$1 (Fig. \ref{fig_1}{\bf a}), the symmetry of the effective potential implies the almost perfect degeneracy of the orbitals $d_{xz}$ and $d_{yz}$ as defined in Fig. \ref{fig_1}{\bf d}. It is this orbital degeneracy that drives the orbital 2CK effect in IrO$_2$ where the $d_{xz}$ and $d_{yz}$ form a local pseudospin basis, while the spin-degenerate conduction electrons act as two independent screening channels. Group theoretical arguments ensure that the exchange scattering processes between  conduction electrons and pseudospin degree of freedom have a form compatible with the Kondo interaction\cite{Cox.87} (cf. \ref{sec:theory}).
Deviations from  perfect symmetry which act as a  pseudo-magnetic field are expected to become visible at lowest $T$. This explains the deviations from the $\sqrt{T}$ behavior observed below $\sim$\,0.5 K in Fig. \ref{fig_2}.
If the two {\itshape defect electrons} localize at sites $M1$ and $M2$, a two-impurity problem might be expected which could lead to inter-site orbital order between the two {\itshape defect electrons}\cite{Mitchell.12}. The non-symmorphic rutile structure, however,  ensures that the $C_4$ rotation axes centered at the sites $M$1 and $M$2  are not parallel.  This together with the local nature of the decomposition provided in Supplementary Equation (\ref{eq:expansion}) (see \ref{sec:theory}) favor local orbital Kondo screening in line with our observation. These conclusions are further corroborated by demonstrating tunability of the orbital 2CK effect to its 1CK counterpart.

{\bf Experimental signatures of orbital 1CK effect in RuO$_2$ NWs.} RuO$_2$ is also a DNL metal with the same non-symmorphic symmetry group as IrO$_2$ but  weaker SOC. In contrast to IrO$_2$, it lacks time-reversal symmetry\cite{Sun.17,Zhu.19}. Based on the analysis for IrO$_2$, we expect that V$_{\rm{\small O}}$'s in RuO$_2$ will drive an orbital 1CK effect. This is indeed borne out by our transport data on RuO$_2$ NWs. Figure \ref{fig_3}{\bf a} shows the $T$ dependence of the time-averaged Kondo resistivity $\langle \rho_{\rm{\small K}} \rangle$ for NW C, where $\rho_{\rm{\small K}}(T) = \rho(T) - \rho_{\rm{\small B0}}$, and $\langle \ldots \rangle$ denotes averaging. (RuO$_2$ NWs often demonstrate temporal $\rho$ fluctuations. Details can be found in \ref{sec:S-rhoExtraction}.) At low $T$, $\langle \rho_{\rm{\small K}} \rangle$ follows the 1CK form\cite{Costi.00}. The inset demonstrates the recovery of a Fermi-liquid ground state with its characteristic $\langle \rho_{\rm{\small K}} \rangle \propto T^2$ behavior below $\sim$\,12 K and unambiguously rules out the 3D EEI effect. Figure \ref{fig_3}{\bf b} shows $\rho(T)$ of NW E in $B$ = 0 and 4 T. For clarity, the $B=0$ data (black symbols) are averaged over time, while the $B = 4$ T data (red symbols) are non-averaged to demonstrate the temporal fluctuations of the low-$T$ $\rho(T)$ (Ref. \citeonline{Lien.11}). Note that, apart from the aforementioned much smaller resistance increase as would be predicted by the 3D EEI effect compared with the experimental results in Figs. \ref{fig_3}{\bf a} and {\bf b}, no $\sqrt{T}$ dependence is detected here. In fact, the low-$T$ resistivity anomalies  conform very well to the 1CK scaling form for three decades in $T/T_{\rm{\small K}}$ (Fig. \ref{fig_4}{\bf a}). Thus, the 3D EEI effect can be safely ruled out as the root of the observed low-$T$ resistivity anomalies in RuO$_2$ NWs.

As a further demonstration of the $B$-field independence, we present in Fig. \ref{fig_3}{\bf c} $\rho(T)$ data for NW A in magnetic fields of strength $B$ = 0, 3 and 5 T. With $T_{\rm{\small K}}^{\rm{\small A}}$\,=\,3 K, NW A has the lowest $T_{\rm{\small K}}$ among NWs A to E (Table \ref{table_1}). The data between 50 mK and 10 K, corresponding to $T/T_{\rm{\small K}} = 0.017 \textendash 3.3$, can be well described by the 1CK function (solid curve). The dash-dotted curves depict the magnetoresistance predicted by the spin-$\frac{1}{2}$ Kondo impurity model\cite{Costi.00} with $g\mu_{\rm{\small B}} B/k_{\rm{\small B}} T_{\rm{\small K}}$ = 1.0, 2.0 and 4.1, as indicated, where $k_{\rm{\small B}}$ is the Boltzmann constant. Our experimental data clearly demonstrate $B$ independence, ruling out a magnetic origin of this phenomenon.

We remark on the relation between the residual resistivity $\rho_{\rm{\small B0}}$ and the concentration of orbital Kondo scatterers n$_{\mbox{\tiny V$_{\mbox{\tiny O}}$}}$ extracted from $\rho_{\rm{\small K0}}$, the Kondo contribution to the $\rho(T$\,$\rightarrow$\,0) (see  \ref{sec:S-numberdensity}), for RuO$_2$ NWs.
With the exception of NW B, our data indicate an approximately linear relation between n$_{\mbox{\tiny V$_{\mbox{\tiny O}}$}}$ and $\rho_{\rm{\small B0}}$ (Table \ref{table_1} and Supplementary Fig. \ref{fig:nVO-rB0}). It is not unexpected that the approximately linear relation between $\rho_{\rm{\small B0}}$ and 
n$_{\mbox{\tiny V$_{\mbox{\tiny O}}$}}$ holds for larger impurity concentrations, corresponding to larger values of $\rho_{\rm{\small B0}}$ as all defects, screened dynamic and static defects, contribute to $\rho_{\rm{\small B0}}$.
This relation strongly demonstrates that the low-$T$ resistivity anomalies are indeed due to V$_{\rm{\small O}}$-driven orbital Kondo effect. (We focus on RuO$_2$ NWs because of the larger number of samples with a larger variation of $\rho_{\rm{\small B0}}$ values compared with IrO$_2$ NWs.)

{\bf Comparison of 2CK and 1CK $\rho(T)$ curves.} Figure \ref{fig_4}{\bf a} demonstrates that $\langle \rho_{\rm{\small K}} \rangle / \rho_{\rm{\small K0}}$ for RuO$_2$ NWs follow the universal 1CK scaling over three decades in $T/T_{\rm{\small K}}$ while $T_{\rm{\small K}}$ ranges from 3 K to 80 K! 
To further substantiate the subtle but distinct differences between the $\sqrt{T}$ dependence of the 2CK behavior in IrO$_2$ NWs from the 1CK scaling form, we plot $\langle \rho_{\rm{\small K}} \rangle / \rho_{\rm{\small K0}}$ as a function of $\sqrt{T/T_{\rm{\small K}}}$ for IrO$_2$ NWs A and B1, together with RuO$_2$ NWs B\textendash E and the 1CK function, in Figs. \ref{fig_4}{\bf b} and {\bf c}, respectively. (The value for $\rho_{\rm{\small K0}}$ was identified with the maximum values of the measured $\rho_{\rm{\small K}}(T)$ anomalies.) Figure \ref{fig_4}{\bf d} illustrates that a dilute system of 2CK scattering centers immersed in a metallic host indeed displays a $\sqrt{T}$ term in its low-$T$ $\rho(T)$. This $\sqrt{T}$ power-law behavior is determined by the leading irrelevant operator near the 2CK fixed point\cite{Affleck.93} and captured by the dynamical large-$N$ method\cite{Parcollet.98,Cox.93,Zamani.13}.
\end{results}

\begin{discussion}
Despite the ubiquitous appearance of magnetic Kondo scattering in real quantum materials\cite{Hewson}, no convincing demonstration of the orbital Kondo effect\cite{Kuramoto.16} or the 2CK effect\cite{Moustakas.96,Aleiner.02} exists. Many claims rest on a model of two-level systems immersed in a metallic host as a possible route to 2CK physics\cite{Vladar.83a,Cox.98}. Theoretical arguments have, however, made it clear that this is not a viable route to nonmagnetic Kondo scattering\cite{Moustakas.96,Aleiner.02}.
Moreover, the creation of scattering centers in a real quantum material necessarily places the system in the weakly disordered regime where  a conductance anomaly, the Altshuler-Aronov correction, occurs whose $T$ dependence can be mistaken for a 2CK signature, see, e.g., Refs.\ \citeonline{Zhu.16} and \citeonline{Zhu.17} or Refs. \citeonline{Cichorek.16} and \citeonline{Gnida.17}. Dilution studies on common Kondo lattice systems\cite{Nicklas.12,Yamane.18}, on the other hand, typically create disorder distributions of Kondo temperatures that may result in a behavior of observables, which can easily be mistaken for that of a generic non-Fermi liquid\cite{Miranda.97}.

We have shown that the low-$T$ resistivity anomaly in the transition metal rutile IrO$_2$ is caused by V$_{\rm{\small O}}$'s, demonstrating key signatures of an orbital 2CK effect and ruling out alternative explanations due to, e.g., the EEI effect. The most convincing argument in favor of 2CK physics would be the demonstration of direct tunability of 2CK physics to 1CK physics upon breaking the channel degeneracy. This is difficult, as the channel degeneracy is protected by time-reversal symmetry. A perhaps less direct, yet complementary, demonstration of this tunability is provided by our results for RuO$_2$ NWs which develop an orbital 1CK effect. In RuO$_2$, the antiferromagnetic order breaks the channel degeneracy. Our analysis also indicates that the underlying symmetries which support the existence of DNLs in the Brillouin zones of  both transition metal rutiles also aid the formation of orbital 2CK and 1CK physics.

Materials condensing in the rutile structure type and its derivatives form an abundant and important class that has helped shaping our understanding of correlated matter. The metal-insulator transition in VO$_2$, e.g., has been known for 60 years\cite{Morin.59}, yet its dynamics  is still not fully understood\cite{Wall.18}.
The demonstration that the non-symmorphic rutile space group supports a V$_{\rm{\small O}}$-driven orbital Kondo effect in $M$O$_2$ holds promise for the realization of novel states of matter. The potential richness of orbital Kondo physics, e.g., on superconducting pairing, was recently pointed out in Ref.\ \citeonline{Kuramoto.16} but may be even richer when considering the possibility of its interplay with topological band structures.
Specifically, we envision the creation of a  2CK  non-symmorphic superlattice of V$_{\rm{\small O}}$'s in IrO$_2$ where the 2CK Majorana  modes entangle with the band structure-enforced Dirac excitations forming a  strongly correlated topological non-Fermi liquid state. Understanding its properties will foster deeper insights into the interplay of topology with strong correlations beyond the usual mean field treatment. The theoretical approach to this non-symmorphic superlattice is reminiscent of the topologically garnished strong-coupling fixed-point pioneered in the context of Weyl-Kondo semimetals\cite{Dzsaber.17,Lai.18}, suitably generalized to capture the intermediate coupling physics of the 2CK effect and its low-$T$ excitations. The fabrication of superlattices of Kondo scattering centers has already been demonstrated\cite{Goh.12} while defect engineering of vacancy networks, including V$_{\rm{\small O}}$ networks is currently explored in a range of materials\cite{Simonov.20,Lai.20}.
The specifics of this unique state and its manufacturing are currently being explored.

\end{discussion}

\begin{methods}

\textbf{NW growth.} IrO$_2$ NWs were grown by the metal-organic chemical vapor deposition (MOCVD) method, using (MeCp)Ir(COD) supplied by Strem Chemicals as the source reagent. Both the precursor reservoir and the transport line were controlled in the temperature range of 100\textendash 130$^\circ$C to avoid precursor condensation during the vapor-phase transport. High purity O$_2$, with a flow rate of 100 sccm, was used as the carrier gas and reactive gas. During the deposition, the substrate temperature was kept at $\approx$\,350$^\circ$C and the chamber pressure was held at $\approx$\,17 torr to grow  NWs\cite{Chen.04,Lin.08}. Selected-area electron diffraction (SAED) patterns\cite{Chen.04} and x-ray diffraction (XRD) patterns\cite{Chen.04b} revealed a single-crystalline rutile structure.

RuO$_2$ NWs were grown by the thermal evaporation method based on the vapor-liquid-solid mechanism, with Au nanoparticles as catalyst. A quartz tube was inserted in a furnace. A source material of stoichiometric RuO$_2$ powder (Aldrich, 99.9\%) was placed in the center of the quartz tube and heated to 920\textendash 960$^\circ$C. During the NW growth, an O$_2$ gas was introduced into the quartz tube and the chamber was maintained at a constant pressure of $\approx$\,2 torr. Silicon wafer substrates were loaded at the downstream end of the quartz tube, where the temperature was kept at 450\textendash 670$^\circ$C (Ref. \citeonline{Liu.07}). The morphology and lattice structure of the NWs were studied using XRD and high-resolution transmission electron microscopy (HR-TEM). The XRD patterns demonstrated a rutile structure\cite{Liu.07}, and the HR-TEM images revealed a polycrystalline lattice structure\cite{Yeh.17}.

\noindent\textbf{Electrical measurements.} Submicron Cr/Au (10/100 nm) electrodes for 4-probe $\rho (T)$ measurements were fabricated by the standard electron-beam lithography technique. The electrode fabrication was done after the thermal treatment (annealing and/or oxygenation) of each NW was completed. To avoid electron overheating, the condition for equilibrium, e$V_{\rm{\small s}} \ll k_{\rm{\small B}}T$, was assured in all resistance measurements\cite{Huang.07}, where e is the electronic charge, and $V_{\rm{\small s}}$ is the applied voltage across the energy relaxation length. The electrical-transport measurements were performed on a BlueFors LD-400 dilution refrigerator equipped with room-temperature and low-temperature low-pass filters. The electron temperature was calibrated down to $\lesssim$ 50 mK. In several cases (RuO$_2$ NWs B\textendash E), the measurements were performed on an Oxford Heliox $^3$He cryostat with a base temperature of $\simeq$\,250 mK. The magnetic fields were supplied by superconducting magnets and applied perpendicular to the NW axis in all cases.
\end{methods}
\begin{addendum}
\item The authors are grateful to F.\ R.\ Chen, J.\ J.\ Kai and the late Y.\ S.\ Huang for growing RuO$_2$ and IrO$_2$ NWs, S.\ P.\ Chiu for experimental assistance, T.\ A.\ Costi for providing the 1CK scaling curve from NRG calculations, and Q.\ Si, A.\ M.\ Chang, C.\ H.\ Chung and S.\ Wirth for helpful discussions. Figure 1\textbf{a} was produced with the help of VESTA\cite{Momma.11}. This work was supported by Ministry of Science and Technology, Taiwan (grant Nos. MOST 106-2112-M-009-007-MY4, 108-3017-F-009-004, and 108-2811-M-009-500) and the Center for Emergent Functional Matter Science of National Chiao Tung University from The Featured Areas Research Center Program within the framework of the Higher Education Sprout Project by the Ministry of Education (MOE) in Taiwan. F.\ Z.\ and J.\ K.\ acknowledge financial support by the Deutsche
Forschungsgemeinschaft (DFG) through SFB/TR 185 (277625399) and the Cluster of Excellence ML4Q (390534769). Work at Zhejiang University  was in part supported by the National Key R\&D Program of the MOST of China, grant No. 2016YFA0300202 and the National Science Foundation of China, grant No. 11774307.

\item[Author contributions] S.\ S.\ Y.\ , S.\ K.\  and J.\ J.\ L.\ conceived the experiment. S.\ S.\ Y.\ and A.\ S.\ L.\  carried out electrical-transport measurements. T.\ K.\ S.\ , C.\ C.\ L.\ and A.\ S.\ L.\  fabricated 4-probe NW devices with thermal treatments. F.\ Z.\  performed dynamical large-$N$ calculations. J.\ K.\  provided theoretical support. S.\ S.\ Y.\ , S.\ K.\  and J.\ J.\  L.\ analyzed and explained the data, and wrote the manuscript.
\item[Competing interests] The authors declare no competing interests.
\item[Correspondence] Correspondence and requests for materials should be addressed to Stefan Kirchner (email: stefan.kirchner@correlated-matter.com) or Juhn-Jong Lin (email: jjlin@mail.nctu.edu.tw).
\item[Data availability] All data collected or analysed during this study is available in the main text or the supplementary information material.
\item[Code availability] Details on the numerics is available upon request from the authors.
\end{addendum}


\begin{figure}
	\centering
	\includegraphics[width=1.0 \linewidth]{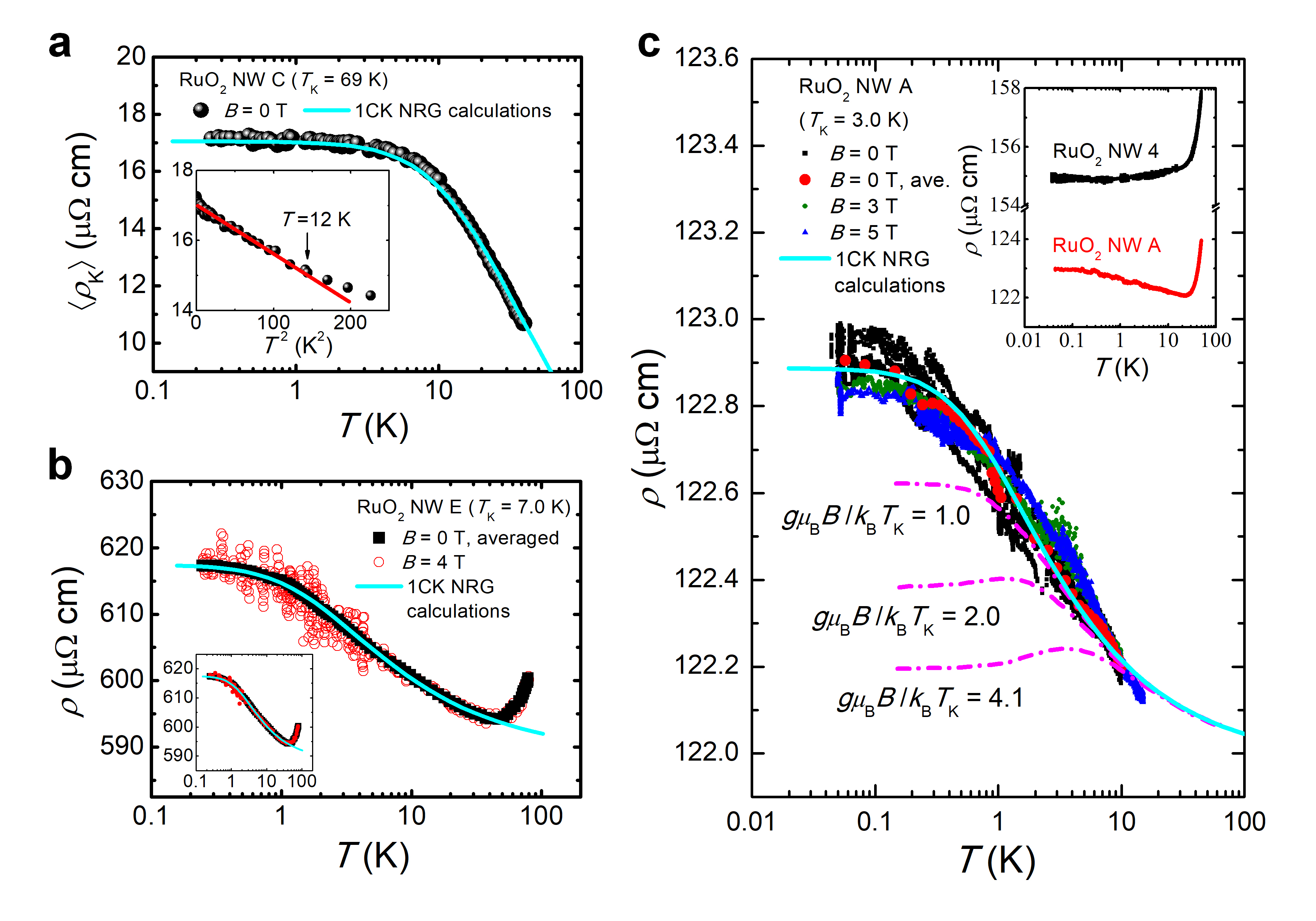}
	\caption{ \textbf{Orbital 1CK resistivity of RuO$_2$ NWs.}
		\textbf{a} Time-averaged Kondo resistivity $\langle \rho_{\rm{\small K}} \rangle$ versus $\log T$ for NW C. The straight line in the inset, which shows a low-$T$ zoom-in, is a guide to the eye. \textbf{b} $\rho$ versus $\log T$ in $B$ = 0 and 4 T for NW E. For clarity, the $B=0$ data are time-averaged, while the 4-T data are non-averaged to demonstrate the temporal resistivity fluctuations at low $T$. The inset shows the time-averaged $B=4$ T data (red symbols), which closely overlap the $B=0$ data. \textbf{c} $\rho$ versus $\log T$ for NW A in $B$ = 0, 3 and 5 T. Occasional resistivity jumps, or random telegraph noise, are observed. The dash-dotted curves depict the magnetoresistance predicted by the spin-$\frac12$ Kondo impurity model (see text). Note that the experimental data are independent of $B$. Inset: Low-$T$ $\rho(T)$ curves of NW A and a reference, oxygenated NW 4 ($d=150$ nm, $\rho$(300 K) = 336 $\upmu\Omega$ cm). In \textbf{a}\textendash \textbf{c}, the solid curve shows the $B=0$ numerical renormalization group result for 1CK effect\cite{Costi.94}.}
	\label{fig_3}
\end{figure}

\begin{figure}
	\centering
	\includegraphics[width=1.0\linewidth]{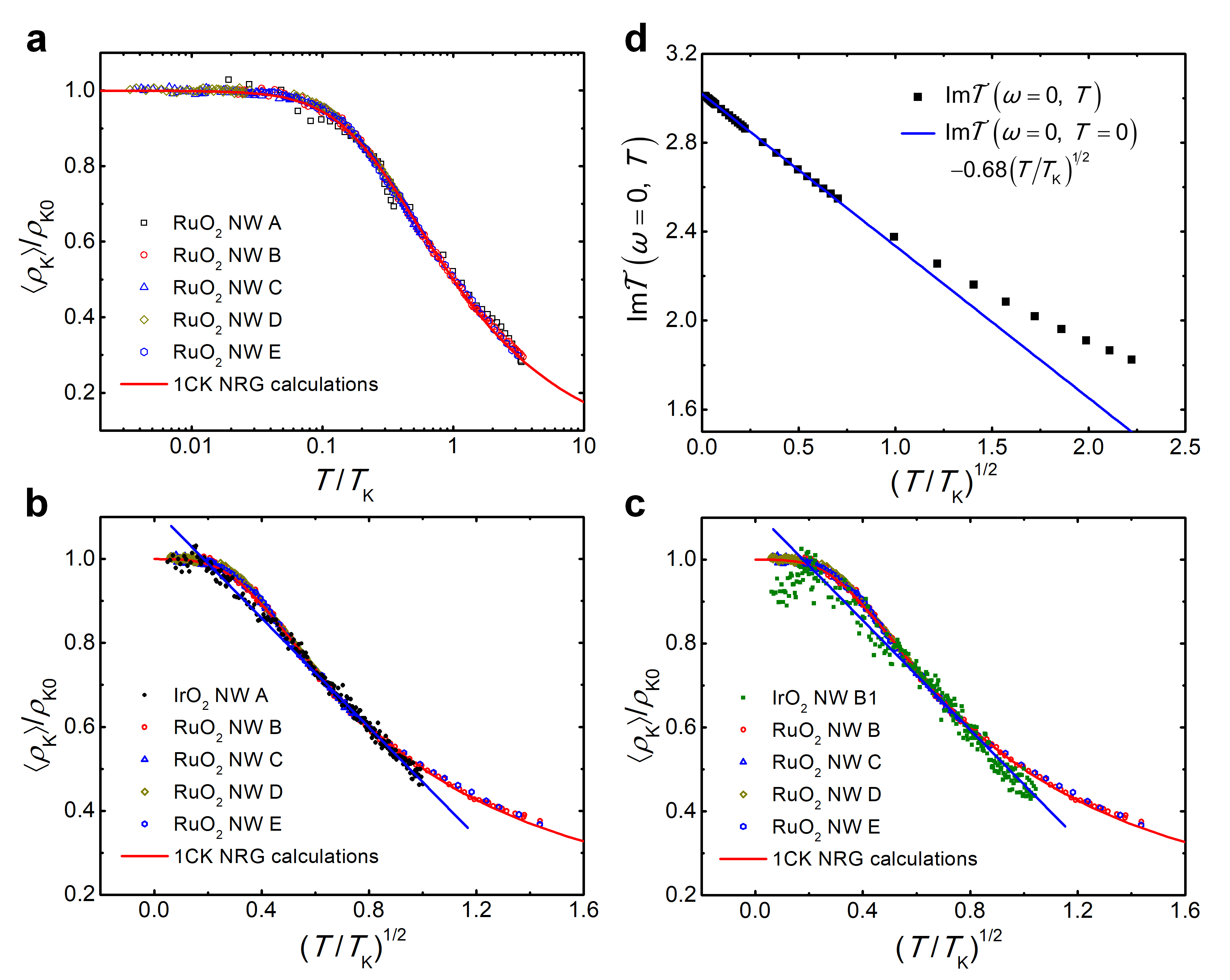}
	\caption{\textbf{Comparison of 2CK and 1CK resistivities.}
		\textbf{a} Normalized Kondo resistivity $\langle \rho_{\rm{\small K}} \rangle / \rho_{\rm{\small K0}}$ versus $T/T_{\rm{\small K}}$ for RuO$_2$ NWs A\textendash E manifests the 1CK scaling form (solid curve) for over three decades of reduced temperature. \textbf{b} $\langle \rho_{\rm{\small K}} \rangle / \rho_{\rm{\small K0}}$ versus $\sqrt{T/T_{\rm{\small K}}}$ for IrO$_2$ NW A and RuO$_2$ NWs B\textendash E. The data of IrO$_2$ NW A obeys a $\sqrt{T}$ law between 0.39 and 21 K. For clarity, the experimental data points for RuO$_2$ NWs are plotted with small open symbols. \textbf{c} $\langle \rho_{\rm{\small  K}} \rangle / \rho_{\rm{\small K0}}$ of IrO$_2$ NW B1 obeys a $\sqrt{T/T_{\rm{\small K}}}$ law between 0.66 and 22 K, distinctively deviating from the 1CK function. \textbf{d} Results for the resistivity of a diluted system of 2CK impurities in a metallic host evaluated using a dynamical large-$N$ limit (black symbols), which follows a $\sqrt{T/T_{\rm{\small K}}}$ law at low $T$ (see text and  \ref{sec-SlargeN}). The ordinate is plotted in unit of half-bandwidth $W$ = 4 eV (Ref. \citeonline{deAlmeida.06}).}
	\label{fig_4}
\end{figure}

\newpage
\noindent\textbf{Table 1}

\renewcommand{\arraystretch}{0.8}
\begin{table*}[h!]
	\centering
	\begin{tabular}{|lccccccccc|}
		\toprule
		\rowcolor{gray!50}
		{\rm NW}%
		&$d$%
		&$\rho(300\,{\rm K})$%
		&$\rho_{\rm{\small B0}}$%
		&$\ell(10\,{\rm K)}$%
		&$D(10\,{\rm K})$%
		&$\rho_{\rm{\small K0}}$%
		&$T_{\rm{\small K}}$%
		&n$_{\mbox{\tiny V$_{\mbox{\tiny O}}$}}$%
		&n$_{\mbox{\tiny V$_{\mbox{\tiny O}}$}}$/n$_{\rm{\small O}}\,(\%)$%
		\\
		\midrule
		{\rm IrO$_2$ A}  &$130$ &$147$ &$109$  &$2.5$ &$4.2$ &$(0.65)$ &$(20)$  &$\sim$\,1.9$\times$$10^{25}$ &$\sim 0.031$ \\
		\midrule
		{\rm IrO$_2$ B1} &$190$ &$104$ &$73.9$ &$3.7$ &$6.2$  &$(0.72)$ &$(20)$  &$\sim$\,2.2$\times$$10^{25}$ &$\sim 0.036$ \\
		\midrule
		{\rm IrO$_2$ B2} &$190$ &$106$ &$75.0$ &$3.6$ &$6.0$  &$(0.45)$ &$(20)$  &$\sim$\,1.4$\times$$10^{25}$ &$\sim 0.023$ \\
		\midrule \midrule
		{\rm RuO$_2$ A}  &$53$  &$193$ &$122$  &$2.2$ &$4.0$  &$0.94$   &$3.0$   &$\sim$\,1.5$\times$$10^{25}$ &$\sim 0.025$ \\
		\midrule
		{\rm RuO$_2$ B}  &$67$  &$163$ &$120$  &$2.3$ &$4.2$  &$14$     &$12$    &$\sim$\,2.3$\times$$10^{26}$ &$\sim 0.38$ \\
		\midrule
		{\rm RuO$_2$ C}  &$54$  &$589$ &$434$  &$0.63$ &$1.2$ &$17$     &$69$    &$\sim$\,2.7$\times$$10^{26}$ &$\sim 0.44$ \\
		\midrule
		{\rm RuO$_2$ D}  &$120$ &$245$ &$160$  &$1.7$  &$3.1$ &$7.0$    &$80$    &$\sim$\,1.1$\times$$10^{26}$ &$\sim 0.18$ \\
		\midrule
		{\rm RuO$_2$ E}  &$47$  &$761$ &$587$  &$0.47$ &$0.9$ &$30$     &$7.0$   &$\sim$\,4.8$\times$$10^{26}$ &$\sim 0.79$ \\
		\bottomrule
	\end{tabular}
	\caption{\label{table_1}%
		\textbf{Relevant parameters for $M$O$_2$ NWs.} Diameter $d$ is in nm, room-temperature resistivity $\rho$(300\,K), residual resistivity $\rho_{\rm{\small B0}}$, and Kondo resistivity in the unitary limit $\rho_{\rm{\small K0}}$ are in $\upmu \Omega$ cm, the electron mean free path $\ell$(10\,K) is in nm, the electron diffusion constant $D$(10\,K) is in cm$^2$ s$^{-1}$, the Kondo temperature $T_{\rm{\small K}}$ is in K, and the number density of oxygen vacancies n$_{\mbox{\tiny V$_{\mbox{\tiny O}}$}}$ is in m$^{-3}$. n$_{\rm{\small O}}$ denotes the oxygen atom number density in the rutile structure. In all 4-probe configuration for transport measurements, the length between the two voltage probes is $\sim$\,1 $\upmu$m. The $\ell$(10\,K) and $D=1/[\rho e^2 N(E_{\rm{\small F}})]=\frac13 v_{\rm{\small F}} \ell$ values are calculated through the free-electron model, where $N(E_{\rm{\small F}})$ is the density of states at the Fermi energy, and the Fermi velocity $v_{\rm{\small F}} \approx 5.0\times 10^{5}$ and $5.5 \times 10^{5}$ m s$^{-1}$ in IrO$_2$ and RuO$_2$, respectively. For each IrO$_2$ NW, we have empirically taken the $\rho_{\rm{\small K0}}$ value to be the maximum value of the measured Kondo resistivity at $\sim$\,0.5 K and $T_{\rm K} \simeq 20$ K. These values are listed in parentheses. IrO$_2$ NW B has been measured twice before and after oxygenation in air and labeled B1 (first measurement) and B2 (second measurement).}
\end{table*}


\newpage
\setcounter{page}{1}
\beginsupplement
\makeatletter
\renewcommand*{\@biblabel}[1]{\hfill #1.}
\makeatother
\spacing{1}
\begin{center} 
	\Large {\textbf{SUPPLEMENTARY INFORMATION}}\\[2ex] 

	\textbf{Oxygen vacancy-driven orbital multichannel Kondo effect in Dirac nodal line metals IrO$_2$ and RuO$_2$} 

    \normalsize
    Yeh et al.

	
\end{center}

\newpage

\section{Thermal annealing effect on the number density of oxygen vacancies n$_{\mbox{\tiny V$_{\mbox{\tiny O}}$}}$ in $M$O$_2$ rutiles} 
\label{sec:S-annealing}

\begin{figure}[!ht]
	\begin{center}
		\includegraphics[width=0.45\linewidth]{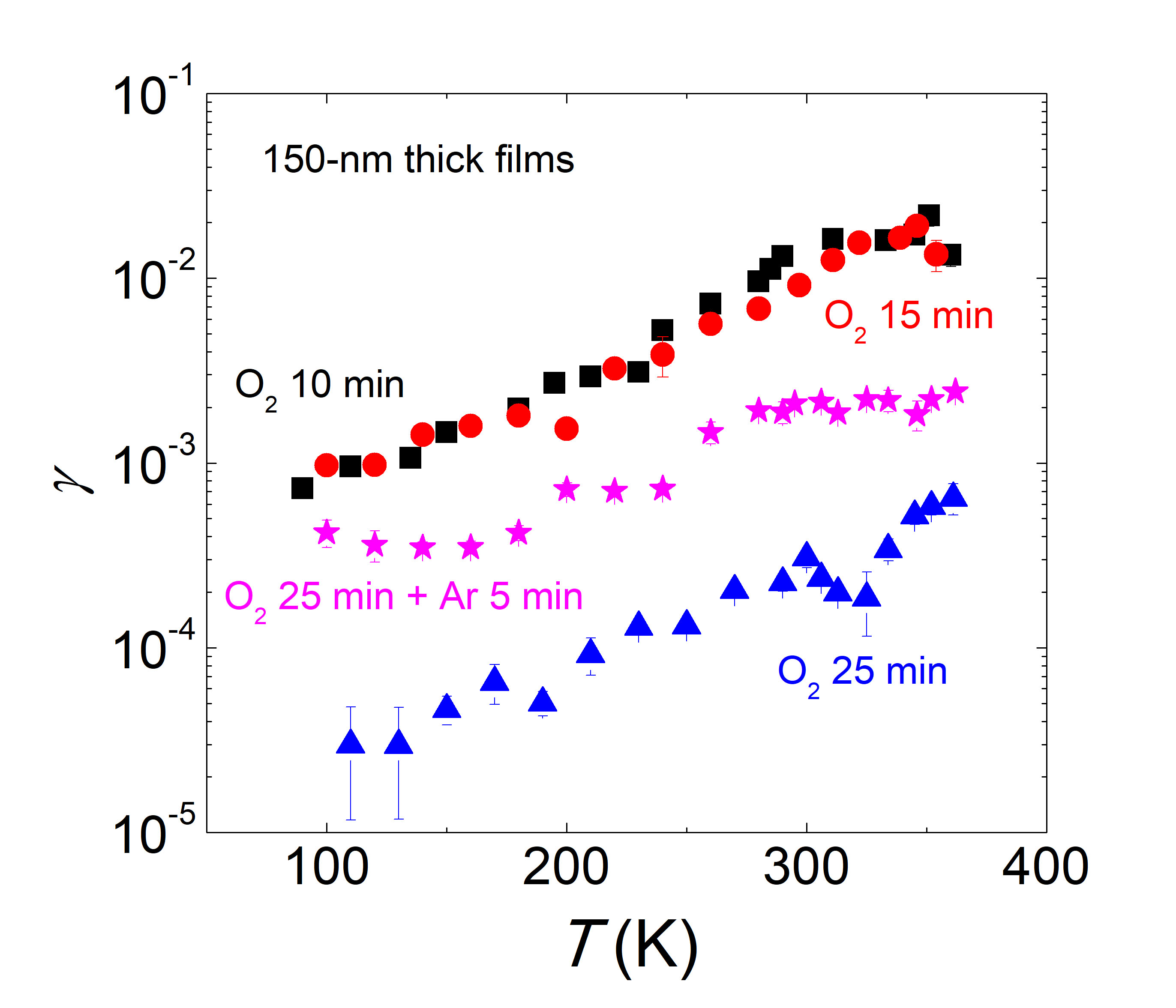}
		\caption{\textbf{Hooge parameter versus temperature for RuO$_2$ films.} Description in the text.}
		\label{annealing}
	\end{center}
	\label{fig_Anneal}
\end{figure}

The number density of oxygen vacancies n$_{\mbox{\tiny V$_{\mbox{\tiny O}}$}}$ in a given sample can be tuned by thermal annealing. The annealing conditions for the $M$O$_2$ NWs studied in this work are listed below:
\begin{itemize}
\setlength\itemsep{0em}
    \item IrO$_2$ NW A was annealed at 300$^\circ$C in vacuum for 1 h.
    \item IrO$_2$ NW B1 was annealed at 450$^\circ$C in vacuum for 10 min.
    \item IrO$_2$ NW B2 was oxygenated in air for $\sim$\,5 months after the first measurement.
    \item IrO$_2$ NW 3 was oxygenated in air.
    \item RuO$_2$ NW A was annealed at 450$^\circ$C in argon for 5 min.
    \item RuO$_2$ NWs B\textendash E were as-grown.
    \item RuO$_2$ NW 4 was oxygenated in air.
\end{itemize}
\vskip -10pt
\noindent \textbf{A control experiment of RuO$_2$ films.} In addition to NWs, we have recently grown a series of 150-nm-thick RuO$_2$ films and carried out a control experiment of the annealing effect under various atmospheric conditions to extract the n$_{\mbox{\tiny V$_{\mbox{\tiny O}}$}}$ values\citeS{Yeh.18s}. The n$_{\mbox{\tiny V$_{\mbox{\tiny O}}$}}$ values in the samples have been inferred from both x-ray photoelectron spectroscopy and 1/$f$ noise studies. These two independent methods give consistent results. Our 1/$f$ noise measurements demonstrate the relation $\gamma \propto$ n$_{\mbox{\tiny V$_{\mbox{\tiny O}}$}}$. Here $\gamma$ is the Hooge parameter which characterizes the magnitude of the voltage noise power spectrum density $S_V(f) = \gamma V^2/N_{\rm c}f$, where $V$ is the bias voltage, $f$ is the frequency, and $N_{\rm c}$ is the total number of charge carriers in the sample. Supplementary Fig. \ref{annealing} is reproduced from figure 6(a) of Ref. \citeS{Yeh.18s}, which shows the variation of $\gamma$ with $T$ for four RuO$_2$ films that underwent different annealing conditions (at 500$^\circ$C), as indicated. We see that annealing for 25 min in O$_2$ greatly reduces the $\gamma$ value (blue symbols), whereas an additional 5 min annealing in Ar significantly increases the $\gamma$ value (pink symbols). Thus, annealing in O$_2$ reduces the n$_{\mbox{\tiny V$_{\mbox{\tiny O}}$}}$ value in the sample. In contrast,  annealing in Ar increases the n$_{\mbox{\tiny V$_{\mbox{\tiny O}}$}}$ value in the sample.

\section{Extraction of the anomalous contribution to the resistivity}
\label{sec:S-rhoExtraction}

Our measurements of $\rho(T)$ indicate that above $\sim$\,a few (tens) degrees Kelvin, depending on the NWs, the variation of $\rho$ with $T$ in both IrO$_2$ and RuO$_2$ NWs obeys the Boltzmann transport equation, as previously established\citeS{Lin.04s}. Thus, the residual resistivity $\rho_{\rm{\small B0}}$ (listed in Table \ref{table_1}) and the Debye temperature, $\theta_{\rm D}$, of each NW can be extracted by the standard method.

\begin{figure}
	\centering
	\includegraphics[width=0.85\linewidth]{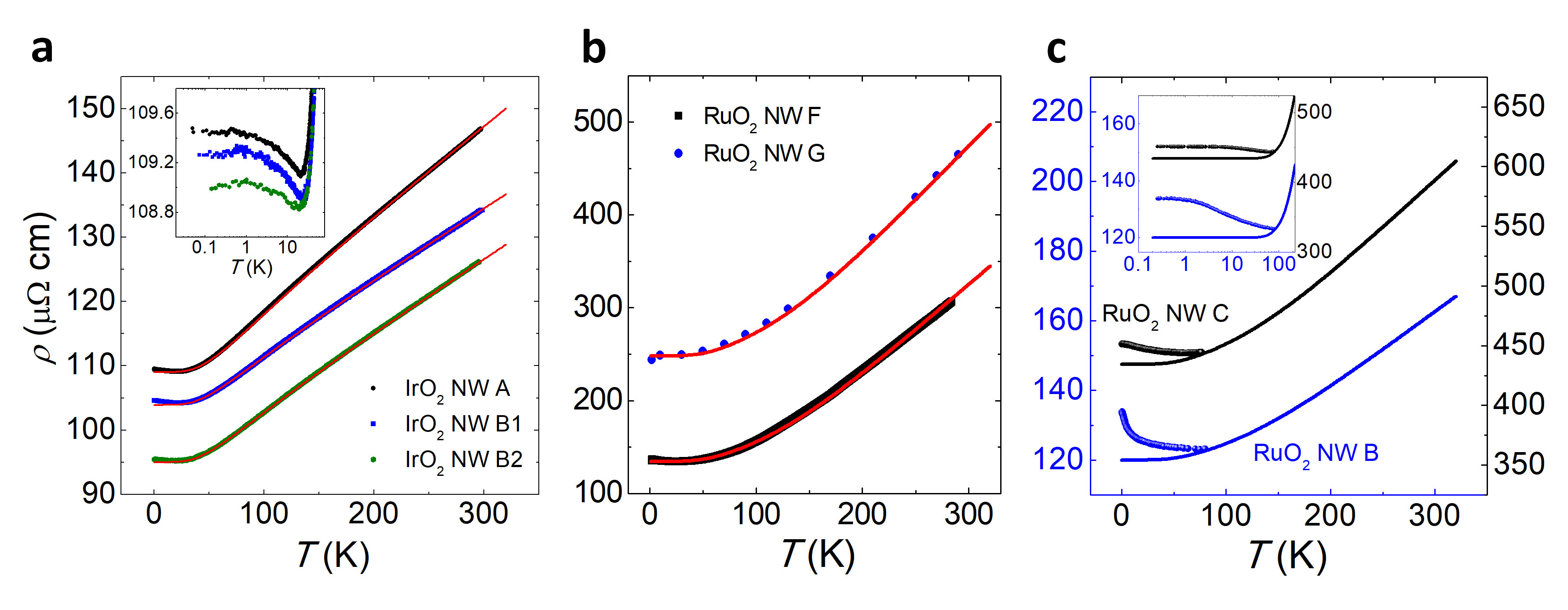}
	\caption{\textbf{Resistivity versus temperature for IrO$_2$ and RuO$_2$ NWs.} Description in the text.}
	\label{fig:restivityFIT}
\end{figure}

Supplementary Fig.\ \ref{fig:restivityFIT}(a) shows $\rho(T)$ data for three IrO$_2$ NWs. The solid curves are Boltzmann transport predictions, with the fitted values $\theta_{\rm D}$ = 275, 260 and 260 K for NWs A, B1 and B2, respectively. For clarity, the data for NWs B1 and B2 are offset by 30 and 20 $\upmu\Omega$ cm, respectively. The inset of Supplementary Fig. \ref{fig:restivityFIT}(a), plotted with a $\log T$ scale, demonstrates that the anomalous low-$T$ (Kondo) resistivity increases with decreasing $T$. The data are offset by 34.7 $\upmu\Omega$ cm (NW B1) and 33.6 $\upmu\Omega$ cm (NW B2) for clarity. 

For RuO$_2$ NWs, an additional term due to the coupling of electrons with optical-mode phonons needs to be included to fully describe $\rho(T)$ curves\citeS{Lin.04s}. The fitted values are $\theta_{\rm D}$ = 375 K and $\theta_{\rm E}$ = 790 K for all NWs, where $k_{\rm{\small B}} \theta_{\rm E} / \hbar$ is the optical-phonon frequency. Supplementary Fig.\ \ref{fig:restivityFIT}(b) shows the $\rho (T)$ curves for NWs F (diameter $d=21$ nm) and G ($d=90$ nm), which were aged in air. The solid curves are the Boltzmann transport predictions. In these two NWs, the low-$T$ anomalies are barely seen. Note that in a number of RuO$_2$ NWs we have observed pronounced temporal resistance fluctuations, associated with an increase in the time-averaged resistivity with decreasing $T$ at low $T$, as previously reported\citeS{Lien.11s}. Thus, the averaged resistivity $\langle \rho(T) \rangle = \frac{1}{\triangle t} \int^{t+\triangle t}_t dt' \rho(T,t')$ is our main focus, i.e., we have averaged the measured resistivities to smooth out any time-dependent fluctuations before extracting the low-$T$ anomalies. The solid curves in the main panel of Supplementary Fig. \ref{fig:restivityFIT}(c) show the Boltzmann transport predictions for RuO$_2$ NWs B and C. The symbols are the measured $\langle \rho(T) \rangle$. The inset shows $\rho(T)$ on a lin-log scale to demonstrate the logarithmic behavior of the Kondo effect for $T$ at and above $T_{\rm{\small K}}$. (We also would like to remark that in IrO$_2$ NWs and in a good number of other metallic NWs, such as ITO NWs\citeS{Chiu.09s} and heavily indium-doped ZnO NWs\citeS{Chiu.13s}, we have not found any noticeable temporal resistance fluctuations at low $T$.)
%

\begin{figure}
	\centering
	\includegraphics[width=0.45\linewidth]{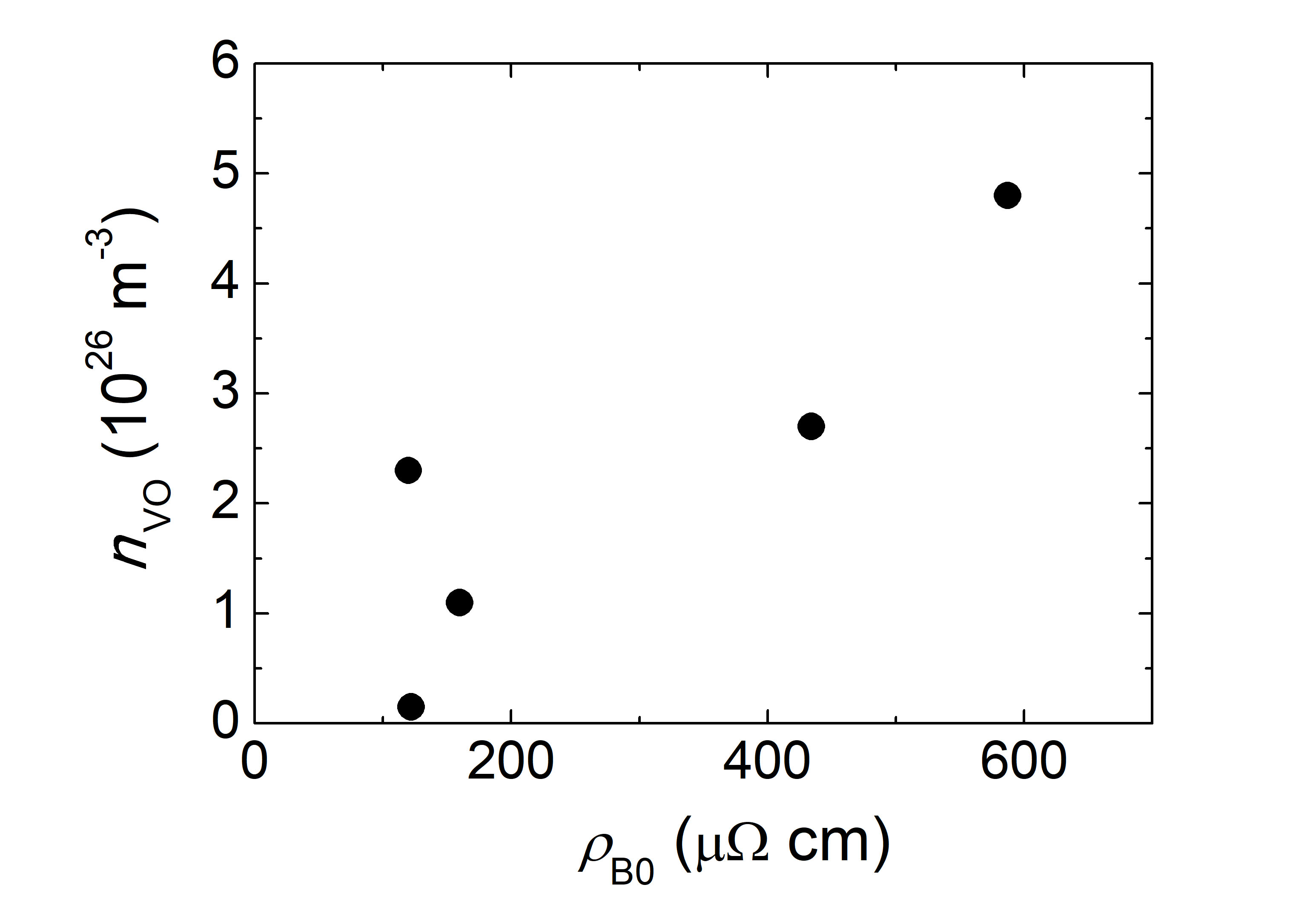}
	\caption{\textbf{Number density of oxygen vacancies (i.e., orbital Kondo scatterers) versus residual resistivity for RuO$_2$ NWs.} Description in the text.}
	\label{fig:nVO-rB0}
\end{figure}

Supplementary Fig. \ref{fig:nVO-rB0} shows  that the relation between our extracted $\rho_{\rm{\small B0}}$ and the concentration of orbital Kondo scatterers n$_{\mbox{\tiny V$_{\mbox{\tiny O}}$}}$, with the exception of NW B, indicates an approximately linear relation for RuO$_2$ NWs (cf. \ref{sec:S-numberdensity} for the extraction of n$_{\mbox{\tiny V$_{\mbox{\tiny O}}$}}$ and Table \ref{table_1} for the extracted values). 

We note that, due to the presence of high levels of point defects in the $M$O$_2$ NWs studied in this work, our measured resistance ratios are small, i.e., $R(300\,{\rm K})/R(10\,{\rm K})<2$, in all samples, implying that the contribution of the electron-phonon scattering to the measured $\rho(T)$ below several tens degrees Kelvin is minute. This specific material property turns out to be a great advantage in the separation of the anomalous resistivity from $\rho_{\rm{\small B0}}$. Quantitatively, the electron-phonon scattering contributes less than a few percent to our measured anomalous resistivity in every NW.

\section{Electron-electron interaction and weak-localization effects}
\label{sec:S-EEI}

Here we expand on the arguments briefly presented in the main text indicating the negligible contribution of EEI effect and weak localization (WL) corrections to the low-$T$ transport properties. In weakly disordered metals, the EEI effect generically dominates over the WL effect in causing characteristic resistivity increases at low $T$, which depends sensitively on the effective dimensionality of the metal\citeS{Altshuler.85s}. For isotropic 3D systems like the $M$O$_2$ rutiles, this EEI correction, $\triangle \rho(T) = \rho(T) - \rho(T_0)$, possesses a $\sqrt{T}$ dependence and is given by\citeS{Lee.85s}
\begin{equation}\label{EEI}
\frac{\triangle \rho(T)}{\rho(T_0)} = - \frac{0.915e^2}{4\pi^2 \hbar} \left( \frac43 - \frac32 \tilde{F} \right) \rho(T_0) \sqrt{\frac{k_{\rm{\small B}}}{\hbar D}}
\left( \sqrt{T} - \sqrt{T_0} \right)\,,
\end{equation}
where $T_0$ is a reference temperature, $e$ is the electronic charge, $2\pi\hbar$ is the Planck constant, $\tilde{F}$ is the screening factor which parameterizes the degree of screening of the EEI effect, and $D$ is the diffusion constant. Empirically, it has been established that, typically, $\tilde{F} \lesssim 0.1$ in weakly disordered metals and alloys\citeS{Akkermans_s,Huang.07s,Lin.87s}. It thus appears as if the EEI correction may offer an alternative explanation for the observed $\sqrt{T}$ transport anomaly found in IrO$_2$ NWs. Indeed, the explanation of a $\sqrt{T}$ resistivity anomaly at low $T$ in terms of a nonmagnetic 2CK effect driven by a dynamical Jahn-Teller effect\citeS{Cichorek.16s} has recently been challenged in favor of an explanation based on the Altshuler-Aronov correction\citeS{Gnida.17s,Cichorek.17s,Gnida.18s}. (The samples studied in Ref. \citeS{Cichorek.16s} were the {\itshape layered} compound ZrAs$_{1.58}$Se$_{0.39}$.) It is thus important to irrefutably connect the behavior observed in IrO$_2$ NWs to the orbital 2CK effect. We therefore stress that the observed resistivity rise at low $T$ cannot be due to the EEI effect and the even smaller WL correction. These two effects would only cause a resistivity correction which would be approximately one order of magnitude smaller than the low-$T$ anomalies observed in our measurements on IrO$_2$ and RuO$_2$ NWs\citeS{Lien.11s}. For example, for the IrO$_2$ NW B1 with $\rho(10\,\rm{K})$ = 74 $\upmu\Omega$ cm, $D \simeq$ 6.2 cm$^2$/s, and by taking $\tilde{F} = 0$, Supplementary Eq. (\ref{EEI}) predicts a largest possible resistance increase of $\triangle \rho/\rho \simeq 2.8 \times 10^{-4}$ as $T$ is cooled from 20 to 1 K. In reality, we have observed a much larger resistance increase of $5.1\times 10^{-3}$ (Fig. \ref{fig_2}). The characteristic thermal diffusion length in the EEI effect is calculated to be $L_{\rm T} = \sqrt{D\hbar/k_{\rm{\small B}}T} = 69/\sqrt{T}$ nm (where $T$ is in K), justifying that the NW is 3D with regard to the EEI effect.

The anomalous resistivities we observe in Figs. \ref{fig_2} and \ref{fig_3} are intrinsic properties of IrO$_2$ and RuO$_2$ rutiles, independent of sample geometry. However, in practice, the resistance of a bulk single crystal (e.g., of a volume $\sim$\,0.1$\times$0.1$\times$0.5 mm$^3$) will be too small for the relative resistance change due to the orbital Kondo effect to be detected unless a large current is applied, which is deemed to cause electron heating. On the other hand, the orbital Kondo resistance in a patterned IrO$_2$ or RuO$_2$ film (e.g., of a volume $\sim$\,0.1$\times$1$\times$10 $\upmu$m$^3$) shall be readily detectable by using a small current to avoid electron overheating.  Experiments on several types of $M$O$_2$ films are in progress.

\noindent \textbf{Ruling out surface defect scattering scenario.} In this work, the low-$T$ resistivity anomaly measured in a given NW is a bulk property of the NW. It originates from the coupling of conduction electrons with those V$_{\rm{\small O}}$'s distributing across the radius of the NW. If the dynamic defect scattering were dominantly taking place near the surface of a NW, one would expect  quasi-two-dimensional (quasi-2D), instead of 3D EEI effect to play a role. The resistance increase due to the quasi-2D EEI effect is given by\citeS{Chiu.13s,Lin.87s} 
\begin{equation}
\label{EEI_2D}
\frac{\triangle R_\Box (T)}{R_\Box (T_0)} = - \frac{e^2}{2\pi^2 \hbar} \left( 1 - \frac43 \tilde{F} \right)
R_\Box (T_0)\, \ln \left( \frac{T}{T_0} \right)\,,
\end{equation}
where $R_\Box$ is the sheet resistance. For concreteness, assume the thickness of an active surface defect layer, if any exists, to be of the order of the electron mean free path $\ell \sim$ 3 nm ($\ll L_{\rm T}$). For a residual resistivity $\rho_{\rm{\small B0}} \sim$ 100 $\upmu\Omega$ cm, $R_\Box = \rho_{\rm{\small B0}}/\ell \sim$ 300 $\Omega$. Then, taking $\tilde{F}=0\textendash 0.1$, Supplementary Eq. (\ref{EEI_2D}) predicts a resistance increase $\triangle R_\Box (T)/R_\Box (T_0) \sim$ 1\% as $T$ is cooled from 20 to 1 K. However, in practice, this hypothetical surface layer must be shunted by the much larger current conduction through the bulk of the NW which has a diameter $d\sim$\,100 nm. Thus, this estimate should be corrected by a factor $\sim \ell/(d/2) \sim 1/20$, leading to a reduced relative sheet resistance change of $1\% \times (1/20) \sim$ 0.05\%. This value is one order of magnitude smaller than what is observed in Fig. \ref{fig_2}. Furthermore, this quasi-2D effect predicts a $\ln T$ dependence, distinct from the measured $\sqrt{T}$ characteristic. The quasi-2D WL effect will also cause a (smaller) $\ln T$ dependence which should be suppressed by a magnetic field of $\sim$\,a few T (Refs. \citeS{Chiu.13s,Lin.87s}). These features are definitely not seen in Fig. \ref{fig_2}. Thus, our measured low-$T$ anomalies cannot be ascribed to a surface defect scattering scenario.

\noindent \textbf{Other scenarios compatible with power-law in temperature behavior.}
A characteristic feature of Kondo physics in metals is the increasing scattering rate as $T$ is lowered in contrast to, e.g., phonon scattering. Processes like dynamical Coulomb blockade and Luttinger liquid behavior are also associated with power-law behavior in $T$. These can be safely ruled out as underlying cause in our IrO$_2$ and RuO$_2$ NWs as  their diameters are much larger than the electron mean free paths in these systems which renders electron transport 3D. Moreover, dynamical Coulomb blockade requires contacts with small tunneling conductances to observe the resulting power-law behavior in $T$ and bias voltage across the junction. The resulting power-law exponent is given by the resistance in units of the quantum of resistance $h/(2e^2)$ (Ref. \citeS{Ingold.92s}). Changes in the diameter of a quasi-1D conductor affect the number of transmission channels coupled to the contact and thus the dynamical Coulomb blockade exponent in contrast to the universal exponent we observe. The 3D nature of our NWs also rules out impurities in Luttinger liquids as a possible cause for the observed power-law behavior. In addition, the $T$ dependence of $\rho$ in this case would have a complicated and impurity concentration dependent structure\citeS{Furusaki.93s} in contrast to our observations.

\noindent \textbf{Classical magnetoresistance.} For a (3D) normal metal placed in a magnetic field, the classical magnetoresistance due to the Lorentz force is given by $\Delta R(B)/R(0)=(\mu B)^2=(B/n_e e \rho_{\rm{\small B0}})^{2}$, where $\mu$ is the mobility. In our NWs with charge carrier concentration $n_{\rm e} \sim 1 \times 10^{28}$ m$^{-3}$ and $\rho_{\rm{\small B0}}\sim 100$ $\upmu\Omega$ cm, $\Delta R(B)/R(0) \sim 4 \times 10^{-7}$ ($3 \times 10^{-5}$) in $B=1$ (9) T, which can be completely ignored. For comparison, in IrO$_2$ and RuO$_2$ bulk single crystals\citeS{Lin.04s} with small $\rho_{\rm{\small B0}} \lesssim 1$ $\upmu\Omega$ cm, the classical magnetoresistance is pronounced.


\section{Atomic arrangement around an oxygen vacancy in $M$O$_2$ rutiles}
\label{sec:S-defectStructure}

Here we provide detailed information for the lattice parameters around the oxygen vacancy V$_{\rm O1}$ depicted in Fig. \ref{fig_1}. The distances between neighboring ions and the relevant angles surrounding this representative vacancy are indicated in Supplementary Figs. \ref{fig_Orbital}(a)\textendash (c), with their respective values listed in Table \ref{table_orbital}. (a) The three $M$1, $M$2 and $M$3 ions surrounding V$_{\rm O1}$ form a (planar) isosceles triangle. (b) The four oxygen ions, labeled O4$^\prime$, O1$^\prime$, O5$^\prime$ and O3$^\prime$, surrounding the $M$3 ion form a (planar) rectangle. (c) The four oxygen ions, labeled O4, O9, O10 and O8, surrounding the $M$2 ion form an almost perfect (planar) square. That is, $r_1$ differs slightly from $r_2$, and $\theta_5$ and $\theta_6$ differ slightly from 90$^\circ$. These values are calculated from  density functional theory and given in Refs. \citeS{Persson.IrOs,Persson.RuOs}.\\
%
\begin{figure}
	\centering
	\includegraphics[width=0.85\linewidth]{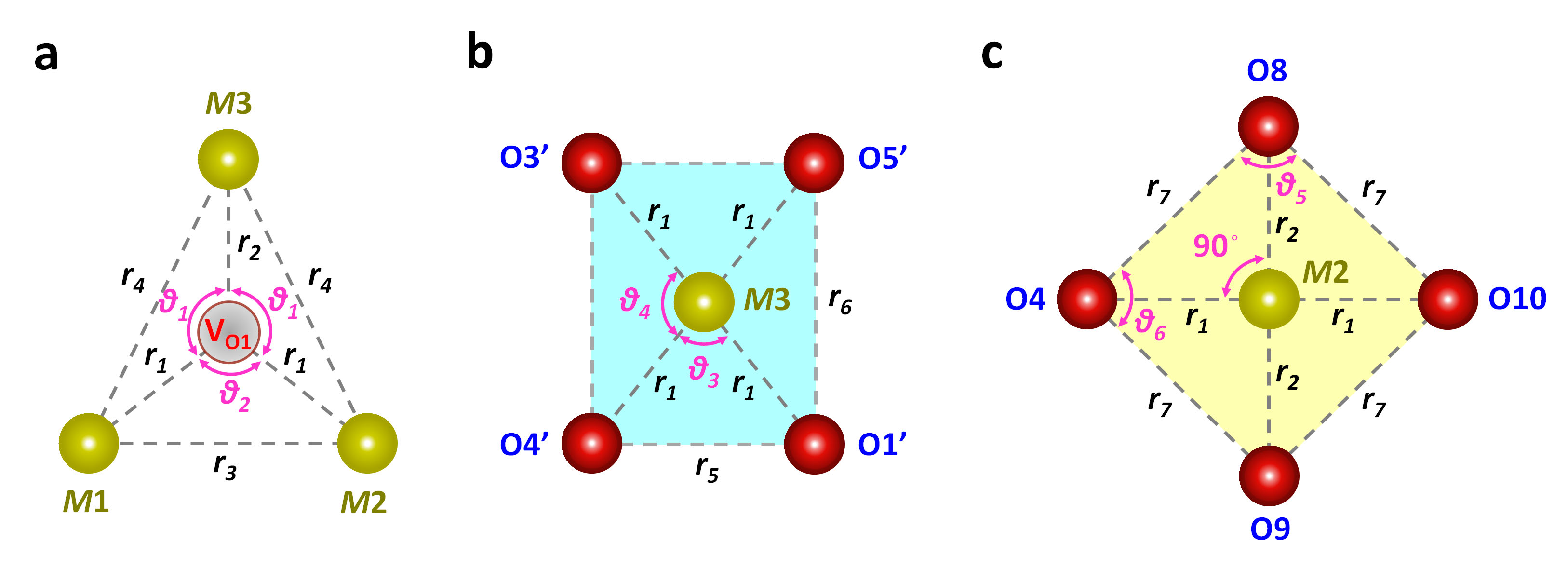}
	\caption{{\bf Vicinity of an oxygen vacancy.} (a) Transition metal ions $M1$, $M2$, and $M3$ in the immediate vicinity of an oxygen vacancy $V_{\rm O1}$. (b) Oxygen positions around the $M3$ site forms a rectangle. (c) The $M2$ site experiences an almost perfect $C_4$ symmetry despite the vicinity to $V_{\rm O1}$. Numerical values for the lattice parameters of IrO$_2$ and RuO$_2$ are provided in Table \ref{table_orbital}.}
	\label{fig_Orbital}
\end{figure}

\begin{table*}[h!]
\renewcommand{\arraystretch}{0.8}
\tabcolsep=4pt
%
	\centering
	\begin{tabular}{|lccccccccccccc|}
		\toprule
		\rowcolor{gray!50}
		$M$O$_2$%
		&$r_1$%
		&$r_2$%
		&$r_3$%
		&$r_4$%
		&$r_5$%
		&$r_6$%
		&$r_7$%
		&$\theta_1$%
		&$\theta_2$%
		&$\theta_3$%
		&$\theta_4$%
		&$\theta_5$%
		&$\theta_6$%
		\\
		\midrule
		{\rm IrO$_2$}    &$2.015$ &$1.982$ &$3.190$ &$3.588$ &$2.464$ &$3.190$ &$2.826$ &$127.7$ &$104.6$ &$75.37$ &$104.6$ &$90.95$ &$89.05$\\
		\midrule
		{\rm RuO$_2$}    &$2.006$ &$1.964$ &$3.140$ &$3.576$ &$2.497$ &$3.140$ &$2.807$ &$128.5$ &$103.0$ &$76.99$ &$103.0$ &$91.20$ &$88.80$\\
		\bottomrule
	\end{tabular}
	\caption{\label{table_orbital}%
		\textbf{Lattice parameters around the oxygen vacancy V$_{\rm O1}$.} The interionic distances $r_i$ ($i=1,...,7$) are in \AA, and the angles $\theta_i$ ($i=1,...,6$) are in degree.}
\end{table*}

\section{Vacancy driven 2CK and 1CK effects and dynamical large-$N$ calculations}
\label{sec:theory}
This section expands the discussion  of orbital 1CK and 2CK physics in the DNL metals RuO$_2$ and IrO$_2$ due to V$_{\rm{\small O}}$'s in the rutile structure.

\subsection{Vacancy driven orbital 2CK and 1CK effects.}
Our starting point is that each V$_{\rm{\small O}}$ will generate two {\itshape defect electrons}. We label the $M$ ion sites closest to the oxygen vacancy  $M1$, $M2$, and $M3$, see Supplementary Fig.~\ref{fig_Orbital}(a). In order to minimize Coulomb repulsion, the defect electrons localize at different $Mi$ near V$_{\rm{\small O}}$. Model studies of V$_{\rm{\small O}}$ in TiO$_2$ are in line with such a picture\citeS{Morgan.10s,Lin.15s,Lechermann.17s}. A small degree of delocalization of the {\itshape defect electron} is acceptable and comparable to the behavior of the standard Anderson impurity model near its Kondo limit. For the sites $M1$, $M2$ and $M3$, the symmetry is reduced due to proximity to V$_{\rm{\small O}}$ as illustrated in Fig.~\ref{fig_1}. In contrast to the {\itshape defect electron} localizing at the $M3$ site, the electron at either the $M2$ or $M1$ site will still experience an almost perfect $C_{4\nu}$ symmetry, see Supplementary Figs.~\ref{fig_Orbital}(b) and \ref{fig_Orbital}(c). (For concreteness, we focus on the $M2$ site.)
The $C_{4\nu}$ group possesses a two-dimensional irreducible representation.

We refer to the corresponding basis functions as $|\Gamma_{\rm{\small 2dim}} 1\rangle$ and  $|\Gamma_{\rm{\small 2dim}} 2\rangle$ which  can be taken as the components of a pseudospin variable. For the $5d$ electrons of IrO$_2$ (and $4d$ electrons of RuO$_2$), the $C_{4\nu}$ at $M2$  ensures the degeneracy  of $d_{xz}$ and $d_{yz}$, where the symmetry axis is parallel to the $\hat{z}$ axis. In the rutile structure, the oxygen ions above and below the square centered around $M2$, i.e., along the $\hat{z}$-axis, are minimally (less than half of their ionic radius) set off from the $C_4$ rotation axis. A V$_{\rm{\small O}}$ thus {\itshape enhances} the symmetry around the $M2$ site, see Fig.~\ref{fig_1}. An orthogonality theorem ensures  that the conduction electrons wavefunction at the $M$2, $|\Psi \rangle$, can be decomposed in terms of the local basis functions as
\begin{equation}
\label{eq:expansion}
|\Psi \rangle =\sum_{\Gamma_{n'}} \sum_{j'}  f^{(\Gamma_{n'} )}_{j'} | \Gamma_{n'} j' \rangle \,,
\end{equation}
where $ | \Gamma_m i\rangle $ denotes the $i^{\rm{\small th}}$ basis function of the $m^{\rm{\small th}}$ irreducible representation. The pseudospin associated with the two-dimensional representation is locally conserved.

In IrO$_2$, the valence of Ir is +IV so that the Ir ion at position $M2$ results in a [Xe]5$d^6$ electron configuration. Arguments reminiscent of those used by Cox in the context of UBe$_{13}$ yield an effective Kondo model where the role of the nonmagnetic $\Gamma_{3}$ is replaced by the two-dimensional irreducible representation of C$_{4\nu}$ (Ref. \citeS{Cox.93s}).
\begin{eqnarray}
\label{eq:OKM}
H=\sum_{k\sigma \mu} \epsilon_k c^\dagger_{k,\mu \sigma}c^{}_{k,\mu \sigma}+J_{\rm{\small K}} \sum_{\stackrel{k,k^\prime,\sigma}{\mu,\nu=1,2}}\vec{\tau}_d\cdot \vec{\tau}_c|_{\mu,\nu} c^{\dagger}_{k,\mu,\sigma}c^{}_{k^\prime,\nu,\sigma}\,, \end{eqnarray}
where $\vec{\tau}_d$ and  $\vec{\tau}_c$ are vectors of Pauli matrices for $d$ and conduction electrons in orbital space, and $J_{\rm{\small K}}$ is a pseudospin exchange coupling constant.
The conduction electron spin projection in Supplementary Eq. (\ref{eq:OKM}) distinguishes the two degenerate screening channels which are necessary for obtaining an overscreened Kondo effect. If the degeneracy between the two screening channels is lifted an orbital 1CK effect ensues. 
This situation occurs in  RuO$_2$. RuO$_2$ is an antiferromagnet with a high N\'{e}el temperature\citeS{Zhu.19s,Berlijn.17s}. In RuO$_2$ the electron configuration of the $M2$ Ru ion is [Kr]$4d^{5}$. Band structure calculations show that  the two $t_{2g}$ orbitals are half-filled\citeS{Berlijn.17s}. As a result, spin excitations of the defect electron are gapped and only orbital fluctuations are possible. These results are in line with our model for IrO$_2$:  For $M2$ sites in the vicinity of V$_{\rm{\small O}}$'s, the orbitals $d_{xz}$ and $d_{yz}$, defined with respect to the $\hat{z}$-axis shown in Fig.~\ref{fig_1}{\bf d}, are half-filled and form the ground state for the {\itshape defect electron}. As in IrO$_2$, V$_{\rm{\small O}}$'s in RuO$_2$ trigger an orbital Kondo effect but due to the lack of time-reversal symmetry, an orbital 1CK effect occurs as demonstrated by the scaling plot of Fig. \ref{fig_4}{\bf a}.

The approximate nature of the almost perfect $C_{4\nu}$ symmetry implies that a low-energy scale exists below which the orbital degeneracy is lifted. We identify this scale with the deviations from the 2CK $\sqrt{T}$ behavior observed in IrO$_2$ NWs around $T\sim$\,0.5 K, see Fig.~\ref{fig_2}. This would be less of an issue for RuO$_2$ NWs due to the more localized nature of the $4d$ vs. the $5d$ orbitals, in line with the experimental observation.

In our arguments we focused on the defect electron localizing at site $M2$.
If the second defect electron localizes at site $M1$, one may expect two-impurity Kondo physics to occur. Due to non-symmorphic rutile structure, however, the $\hat{z}$ axis with respect to which we defined $d_{xz}$ and $d_{yz}$ are not parallel. This fact, together with the local nature of the expansion Supplementary Eq. (\ref{eq:expansion}), favors local orbital Kondo screening over an orbital dimer.

\subsection{Dynamical large-$N$ calculations.}
\label{sec-SlargeN}
The $T$ dependence of the transport lifetime $\tau$ in our IrO$_2$ NWs due to orbital 2CK scatterers can be obtained, e.g., via a dynamical large-$N$ approach. We assume that the concentration of V$_{\rm{\small O}}$'s (n$_{\mbox{\tiny V$_{\mbox{\tiny O}}$}}$) is small enough so that inter-vacancy effects can be ignored. The lifetime $\tau$ is related to $T$-matrix $\mathcal{T}(\omega,T)$ via $\tau(k)^{-1}=\frac{2}{\hbar}c_{\rm{\small 2CK}} \rm{Im}$$\langle k |\mathcal{T}$$(\omega,T)|k\rangle$, where $c_{\rm{\small 2CK}}$ is the concentration of spherically symmetric dynamic scattering centers, which we identify with n$_{\mbox{\tiny V$_{\mbox{\tiny O}}$}}$, and $|k \rangle$ is a plane wave state.
\begin{figure}
	\centering
	\includegraphics[width=0.6\linewidth]{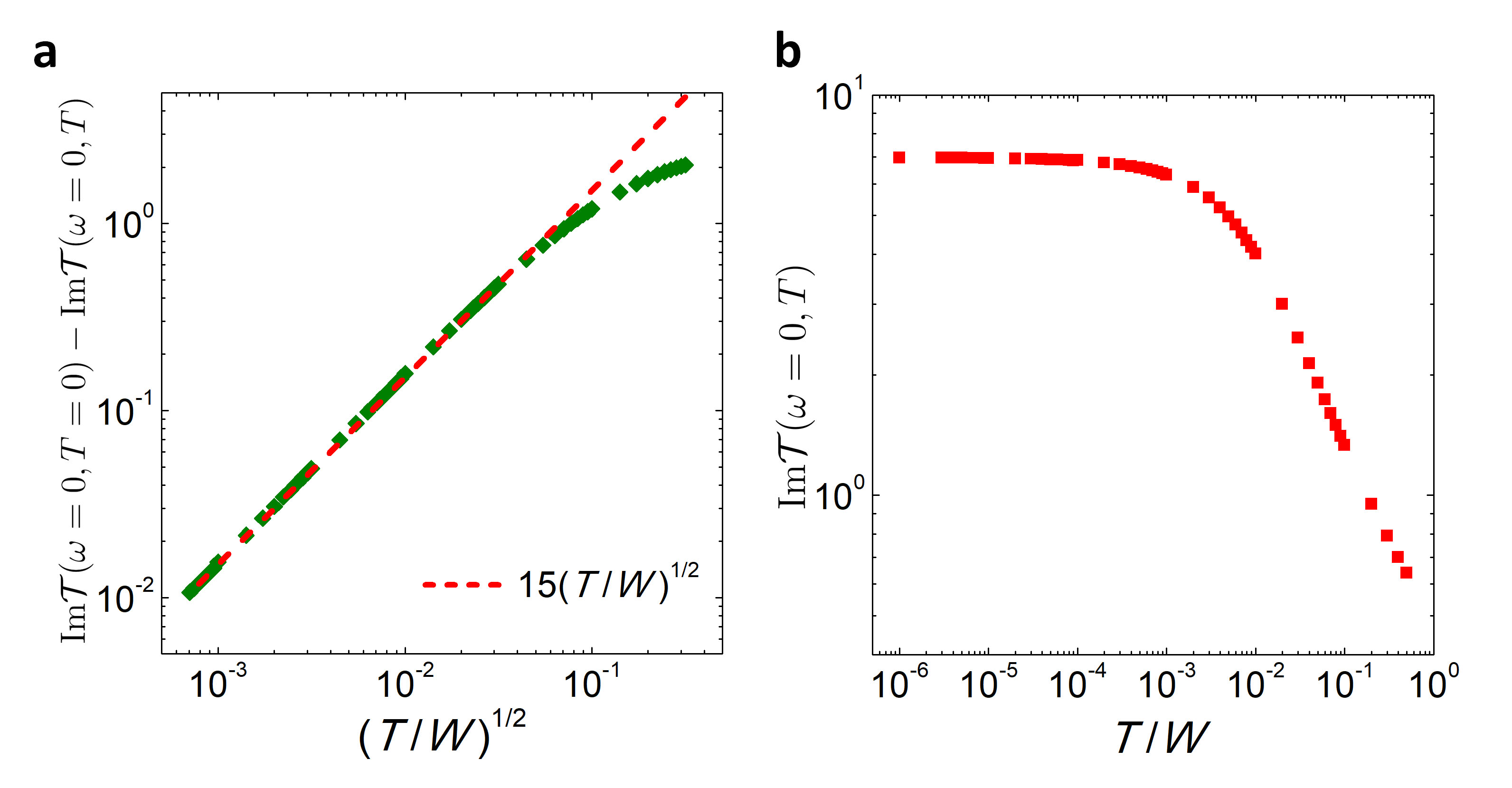}
	\caption{\textbf{Dynamical large-$N$ calculation results.} Imaginary part of the $T$-matrix (in unit of half-bandwidth $W$) for the multi-channel Kondo model within the dynamical large-$N$ limit:
		(a) $\rm{Im}\mathcal{T}$$(\omega=0,T=0)-\rm{Im}\mathcal{T}$$(\omega=0,T)$ versus $\sqrt{T/W}$ for $\kappa=1$. This corresponds to the low-$T$ behavior in IrO$_2$.  
		(b) $\rm{Im}\mathcal{T}$$(\omega=0,T)$ for $\kappa=1/2$ which is equivalent to the one-channel case appropriate for antiferromagnetic RuO$_2$.}
		\label{fig:largeN}
\end{figure}
$\mathcal{T}(\omega,T)$ can be obtained from a dynamical large-$N$ method which is known to give reliable results for multichannel Kondo, including orbital 2CK problems\citeS{Parcollet.98s,Cox.93s,Zamani.13s,Zamani_thesis_s}. In this approach, the symmetry group  is extended to $N$ spin and $M$ orbital channels, resulting in an  SU($N$)$\times$SU($M$) invariant Hamiltonian of the form
\begin{eqnarray}
	\label{eq:Ham}
	H=\sum_{k,\mu,\sigma}\varepsilon_{k}c_{k,\mu,\sigma}^{\dagger}c_{k,\mu,\sigma} 
	 +\frac{J_{\rm{\small K}}}{N}\sum_{\stackrel{k,k'}{\sigma,\mu,\nu}}\Big(f^\dagger_\mu f^{}_\nu-q\delta_{\mu \nu} \Big) c_{k,\mu,\sigma}^{\dagger}c^{}_{k' \nu,\sigma} \,,
\end{eqnarray}
where $\mu, \nu~(1,\ldots,N)$ and $\sigma~(1,\ldots,M)$ are, respectively, the SU$(M)$-channel and SU$(N)$-spin indices and $k, k'$ are momentum indices. The parameter $\kappa$ represents the ration of spin to charge channels. For V$_{\rm{\small O}}$'s in IrO$_2$, $\kappa=1$ due to an equal number of spin and charge channels. If the conduction electron spin degeneracy is lifted, as in RuO$_2$,  $\kappa=1/2$.
Within the large-N approach, one finds
\begin{eqnarray}
\rm{Im} \mathcal{T}(\omega,\it{T})=-\frac{\kappa}{\pi}\int d\epsilon \Big(n_b(\epsilon,\it{T})-n_f(\epsilon+\omega,\it{T})\Big)A_f(\epsilon+\omega,T)A_B(\epsilon,\it{T})\,,
\end{eqnarray}
where $n_{f}$ ($n_{B}$) is the fermionic (bosonic) distribution function, and $A_f$ ($A_B$) is the spectral function associated with the fermionic pseudoparticle representation (bosonic  decoupling field)\citeS{Parcollet.98s,Zamani_thesis_s}. 
For $\kappa=1$, one finds
\begin{eqnarray}
\label{eq:resist}
\rho(T)=\rho(T=0)+A \sqrt{T} \,\,\,\,\,\, {\rm for}\,\,\,\, T \ll T_{\rm{\small K}}\,,
\end{eqnarray}
with $A<0$ as shown in Fig. \ref{fig_4}{\bf d}. The large-$N$ analog of the 1CK case where $\kappa=1/2$ is shown in Supplementary  Fig.~\ref{fig:largeN}(b) which resembles the behavior found in RuO$_2$ NWs. The single-channel case is only semi-quantitatively reproduced by our dynamical large-$N$ approach which does not capture the Fermi-liquid fixed point. We do compare the RuO$_2$ data to NRG calculations by T.~Costi\cite{Costi.94s}.
\section{Extraction of the number density of oxygen vacancies n$_{\mbox{\tiny V$_{\mbox{\tiny O}}$}}$}
\label{sec:S-numberdensity}

The extraction of the n$_{\mbox{\tiny V$_{\mbox{\tiny O}}$}}$ values in our NWs is described below. As discussed, each V$_{\rm{\small O}}$ contributes one \textit{defect electron} to the $M$2 (or $M$1) ion, causing the orbital Kondo effect. Because IrO$_2$ NWs manifest the 2CK effect, the $\rho \propto \sqrt{T}$ characteristic at low $T$ can be compared with the dynamical large-$N$ calculations (Sec. \ref{sec-SlargeN}), which predicts the 2CK resistivity to be 
\begin{equation}
\rho_{\rm 2CK}(T) =\frac{m}{n_e e^2} \frac{ {\rm n}_{\mbox{\tiny V$_{\mbox{\tiny O}}$}}  }{\hbar}\, 2 \Lambda_a\, \rm{Im} \mathcal{T}(\omega=0,\it{T})\,,
\label{eq:rho2CK}
\end{equation}
where $m\sim 1.6\,m_0$ is the effective electron mass ($m_0$ is the free electron mass)\citeS{Choi.06s}, $n_e \sim 1\times 10^{28}$ m$^{-3}$ is the carrier density\citeS{Kawasaki.18s}, $\Lambda_a \sim 1\times 10^{-29}$ m$^{-3}$ is the atomic volume, and $\rm{Im} \mathcal{T}$$(\omega=0,T)$ is the imaginary part of the $T$-matrix calculated from the dynamical large-$N$ method. The n$_{\mbox{\tiny V$_{\mbox{\tiny O}}$}}$ value of every IrO$_2$ NW has been extracted by fitting the measured slope of the $\rho \propto \sqrt{T}$ curve in Fig. \ref{fig_2} to Supplementary Eq. (\ref{eq:rho2CK}). The extracted value is in good agreement with that inferred from the conformal-field-theory calculations\citeS{Affleck.93s}.

As the low-$T$ resistivities of RuO$_2$ NWs conform to the 1CK scaling form, the $\rho_{\rm{\small K0}}$ value in every NW can be inferred. In this unitary limit, the Kondo resistivity is given by\citeS{Vladar.83s}
\begin{equation}
\rho_{\rm{\small K0}} =\frac{m}{n_e e^2} \frac{5 \pi {\rm n}_{\mbox{\tiny V$_{\mbox{\tiny O}}$}}}{12 \hbar g_0}\,,
\label{eq:rhoK0}
\end{equation}
where $m\sim 1.4\,m_0$ (Ref. \citeS{Glassford.93s}), $n_e \sim 1\times 10^{28}$ m$^{-3}$ (Ref. \citeS{Shin.97s}), and the density of states per spin orientation $g_0 \sim 1 \times 10^{47}$ J$^{-1}$ m$^{-3}$ (Ref. \citeS{Glassford.93s}). The n$_{\mbox{\tiny V$_{\mbox{\tiny O}}$}}$ value in every RuO$_2$ NW has been calculated from the extracted $\rho_{\rm{\small K0}}$ value, according to Supplementary Eq. (\ref{eq:rhoK0}).

\section{Kondo temperature distribution}
\label{sec:S-KondoDistr}
Our experimental findings are compatible with a single energy scale, termed Kondo temperature $T_{\rm{\small K}}$, that characterizes the low-$T$ transport in each NW. This implies that the distribution of Kondo temperatures in each NW is sharply peaked around a single $T_{\rm{\small K}}$. This is demonstrated in Supplementary Fig. \ref{fig:exampleFit} where the measured $\langle \rho_{\rm{\small K}}(T) \rangle$ of RuO$_2$ NW C is plotted against NRG results for the $T$-matrix ${\mathcal{T}}(T)$ of the 1CK Kondo model, with the least-squares fitted $T_{\rm{\small K}}=70$ K (solid curve). In addition, we plot ${\mathcal{T}}(T)$ with four selected $T_{\rm{\small K}}$ values (dashed curves) for comparison. Obviously, the theoretical curves with $T_{\rm{\small K}}=35$ K and $T_{\rm{\small K}}=105$ K already deviate significantly from the experimental data. 
%
\begin{figure}[!htb]
	\centering
	\includegraphics[width=0.42\linewidth]{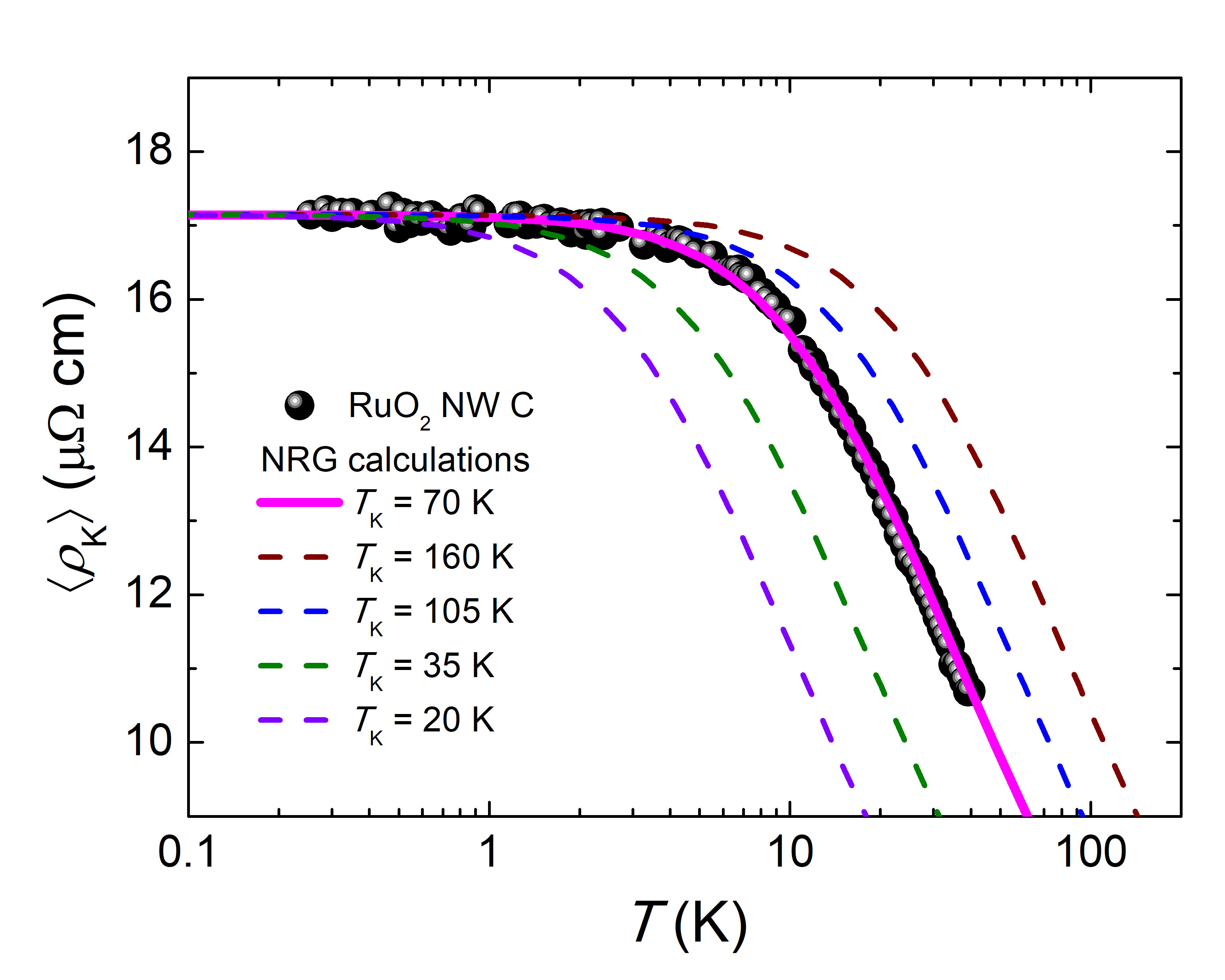}
	\caption{\textbf{Kondo temperature extraction.} NRG calculations for the 1CK model are fitted to $\langle \rho_{\rm{\small K}}(T) \rangle$ of RuO$_2$ NW C with the Kondo temperature $T_{\rm{\small K}}$ as the only adjustable parameter. Solid and dashed curves are characterized by different $T_{\rm{\small K}}$ values, as indicated.}
	\label{fig:exampleFit}
\end{figure}

The Kondo energy scale $T_{\rm{\small K}}$ thus seems to be unaffected by the disorder generated through variations accompanying the oxygen defects. Similar observations have been reported in a variety of systems, including metal nanoconstrictions\citeS{Ralph.92s}, dilute magnetic noble metals\citeS{Mallet.06s}, and other Kondo systems\citeS{Daybell.67s,DiTusa.92s}. This insensitivity of $T_{\rm{\small K}}$ is not entirely unexpected. The exchange coupling constant $J_{\rm{\small K}}$, Supplementary Eq. (\ref{eq:OKM}), reflects the wavefunction overlap between the local basis functions of the dynamic scattering center and the conduction electrons. $J_{\rm{\small K}}$ is therefore not affected by background variations due to nearby V$_{\rm{\small O}}$'s or the presence of the second {\itshape defect electron} on a nearby $M$ ion site. This form of disorder will instead lead to local variations of the background potential. Thus, the situation resembles that considered by Chakravarty and Nayak, who have shown that for weak disorder the distribution of the local density of states is very narrow\citeS{Chakravarty.00s}. Furthermore, the thermodynamics is unaffected near the strong-coupling limit.  While $T_{\rm{\small K}}$ is unaffected by the background variations caused by the weak disorder, it does change from NW to NW. These variations arise from changes in $\rho(E_{\rm F})$ as a result of the stress that exists in each NW, where $\rho(E_{\rm F})$ is the local density of states of conduction electrons coupling to the orbital degree. As the bulk and shear modulus of RuO$_2$ is less than those of IrO$_2$ (Refs. \citeS{Persson.IrOs,Persson.RuOs}), it appears natural that the corresponding variation in $\rho(E_{\rm F})$ is bigger in RuO$_2$ than in IrO$_2$. The observed bigger variations of $T_{\rm{\small K}}$ for different RuO$_2$ NWs compared to those for IrO$_2$ NWs are consistent with this picture.


\begin{thebibliography}{10}
	
	\bibitem{Keimer.15}
	Keimer, B., Kivelson, S.~A., Norman, M.~R., Uchida, S. \& Zaanen, J., From
	quantum matter to high-temperature superconductivity in copper oxides.
	\emph{Nature} \textbf{518}, 179–186 (2015).
	
	\bibitem{Nozieres.80}
	Nozi{\'{e}}res, P. \& Blandin, A., {K}ondo effect in real metals. \emph{J.
		Phys. (Paris)} \textbf{41}, 193--211 (1980).
	
	\bibitem{Vladar.83a}
	Vlad\'{a}r, K. \& Zawadowski, A., Theory of the interaction between electrons
	and the two-level system in amorphous metals. {I}. {N}oncommutative model
	{H}amiltonian and scaling of first order. \emph{Phys. Rev. B} \textbf{28},
	1564--1581 (1983).
	
	\bibitem{Cox.98}
	Cox, D.~L. \& Zawadowski, A., Exotic {K}ondo effects in metals: magnetic ions
	in a crystalline electric field and tunnelling centres. \emph{Adv. Phys.}
	\textbf{47}, 599--942 (1998).
	
	\bibitem{Potok.07}
	Potok, R.~M., Rau, I.~G., Shtrikman, H., Oreg, Y. \& Goldhaber-Gordon, D.,
	Observation of the two-channel {K}ondo effect. \emph{Nature} \textbf{446},
	167--171 (2007).
	
	\bibitem{Keller.15}
	Keller, A.~J., Peeters, L., Moca, C.~P., Weymann, I., Mahalu, D., Umansky, V.,
	Zar\'{a}nd, G. \& Goldhaber-Gordon, D., Universal {F}ermi liquid crossover
	and quantum criticality in a mesoscopic system. \emph{Nature} \textbf{526},
	237 (2015).
	
	\bibitem{Iftikhar.15}
	Iftikhar, Z., Jezouin, S., Anthore, A., Gennser, U., Parmentier, F.~D.,
	Cavanna, A. \& Pierre, F., Two-channel {K}ondo effect and renormalization
	flow with macroscopic quantum charge states. \emph{Nature} \textbf{526}, 233
	(2015).
	
	\bibitem{Iftikhar.18}
	Iftikhar, Z., Anthore, A., Mitchell, A.~K., Parmentier, F.~D., Gennser, U.,
	Ouerghi, A., Cavanna, A., C.~Mora~and, P.~S. \& Pierre, F., Tunable quantum
	criticality and super-ballistic transport in a “charge” {K}ondo circuit.
	\emph{Science} \textbf{360}, 1315 (2018).
	
	\bibitem{Burkov.16}
	Burkov, A.~A., Topological semimetals. \emph{Nat. Mater.} \textbf{15},
	1145--1148 (2016).
	
	\bibitem{Yang.18}
	Yang, S.-Y., Yang, H., Derunova, E., Parkin, S. S.~P., Yan, B. \& Ali, M.~N.,
	Symmetry demanded topological nodal-line materials. \emph{Adv. Phys.: X}
	\textbf{3}, 1414631 (2018).
	
	\bibitem{Dzsaber.17}
	Dzsaber, S., Prochaska, L., Sidorenko, A., Eguchi, G., Svagera, R., Waas, M.,
	Prokofiev, A., Si, Q. \& Paschen, S., Kondo Insulator to Semimetal
	Transformation Tuned by Spin-Orbit Coupling. \emph{Phys. Rev. Lett.}
	\textbf{118}, 246601 (2017).
	
	\bibitem{Ourmazd.87}
	Ourmazd, A. \& Spence, J. C.~H., Detection of oxygen ordering in
	superconducting cuprates. \emph{Nature} \textbf{329}, 425–427 (1987).
	
	\bibitem{Lin.13}
	Lin, C. \& Demkov, A.~A., Electron Correlation in Oxygen Vacancy in
	{SrTiO$_3$}. \emph{Phys. Rev. Lett.} \textbf{111}, 217601 (2013).
	
	\bibitem{Lin.15}
	Lin, C., Shin, D. \& Demkova, A.~A., Localized states induced by an oxygen
	vacancy in rutile {TiO$_2$}. \emph{J.\,Appl.\,Phys.} \textbf{117}, 225703
	(2015).
	
	\bibitem{Lechermann.17}
	Lechermann, F., Heckel, W., Kristanovski, O. \& M\"uller, S., Oxygen-vacancy
	driven electron localization and itinerancy in rutile-based
	{${\mathrm{TiO}}_{2}$}. \emph{Phys. Rev. B} \textbf{95}, 195159 (2017).
	
	\bibitem{Jovic.18}
	Jovic, V., Koch, R.~J., Panda, S.~K., Berger, H., Bugnon, P., Magrez, A.,
	Smith, K.~E., Biermann, S., Jozwiak, C., Bostwick, A., Rotenberg, E. \&
	Moser, S., Dirac nodal lines and flat-band surface state in the functional
	oxide {RuO$_2$}. \emph{Phys.\ Rev.\ B} \textbf{98}, 241101 (2018).
	
	\bibitem{Nelson.19}
	Nelson, J.~N., Ruf, J.~P., Lee, Y., Zeledon, C., Kawasaki, J.~K., Moser, S.,
	Jozwiak, C., Rotenberg, E., Bostwick, A., Schlom, D.~G., Shen, K.~M. \&
	Moreschini, L., Dirac nodal lines protected against spin-orbit interaction in
	{IrO$_2$}. \emph{Phys. Rev. Mater.} \textbf{3}, 064205 (2019).
	
	\bibitem{Zhao.16}
	Zhao, Y.~X. \& Schnyder, A.~P., Nonsymmorphic symmetry-required band crossings
	in topological semimetals. \emph{Phys. Rev. B} \textbf{94}, 195109 (2016).
	
	\bibitem{Sun.17}
	Sun, Y., Zhang, Y., Liu, C.-X., Felser, C. \& Yan, B., Dirac nodal lines and
	induced spin {H}all effect in metallic rutile oxides. \emph{Phys. Rev. B}
	\textbf{95}, 235104 (2017).
	
	\bibitem{Coleman.95}
	Coleman, P., Ioffe, L.~B. \& Tsvelik, A.~M., Simple formulation of the
	two-channel {K}ondo model. \emph{Phys. Rev. B} \textbf{52}, 6611--6627
	(1995).
	
	\bibitem{Moustakas.97}
	Moustakas, A.~L. \& Fisher, D.~S., Two-channel {K}ondo physics from tunneling
	impurities with triangular symmetry. \emph{Phys.~Rev.~B} \textbf{55},
	6832--6846 (1997).
	
	\bibitem{Aleiner.02}
	Aleiner, I. \& Controzzi, D., {Nonexistence of a strong coupling two-channel
		{K}ondo fixed point for microscopic models of tunneling centers}. \emph{Phys.
		Rev. B} \textbf{66}, 045107 (2002).
	
	\bibitem{Yeh.18}
	Yeh, S.-S., Gao, K.~H., Wu, T.-L., Su, T.-K. \& Lin, J.-J., Activation Energy
	Distribution of Dynamical Structural Defects in {RuO$_2$} Films. \emph{Phys.
		Rev. Appl.} \textbf{10}, 034004 (2018).
	
	\bibitem{Altshuler.85}
	Altshuler, B.~L. \& Aronov, A.~G., \emph{Electron-electron interaction in
		disordered conductors}, book section~1 (North-Holland Physics Publishing,
	Amsterdam, The Netherlands, 1985).
	
	\bibitem{Lee.85}
	Lee, P.~A. \& Ramakrishnan, T.~V., Disordered electronic systems. \emph{Rev.
		Mod. Phys.} \textbf{57}, 287--337 (1985).
	
	\bibitem{Ping.15}
	Ping, Y., Galli, G. \& Goddard, W.~A., Electronic Structure of {IrO$_2$}: The
	Role of the Metal d Orbitals. \emph{J. Phys. Chem. C} \textbf{119},
	11570--11577 (2015).
	
	\bibitem{Cox.87}
	Cox, D.~L., Quadrupolar {Kondo} effect in uranium heavy-electron materials?
	\emph{Phys. Rev. Lett.} \textbf{59}, 1240--1243 (1987).
	
	\bibitem{Mitchell.12}
	Mitchell, A.~K., Sela, E. \& Logan, D.~E., Two-Channel {K}ondo Physics in
	Two-Impurity {K}ondo Models. \emph{Phys. Rev. Lett.} \textbf{108}, 086405
	(2012).
	
	\bibitem{Zhu.19}
	Zhu, Z.~H., Strempfer, J., Rao, R.~R., Occhialini, C.~A., Pelliciari, J., Choi,
	Y., Kawaguchi, T., You, H., Mitchell, J.~F., Shao-Horn, Y. \& Comin, R.,
	Anomalous Antiferromagnetism in Metallic {RuO$_2$} Determined by Resonant
	{X}-ray Scattering. \emph{Phys.\ Rev.\ Lett.} \textbf{122}, 017202 (2019).
	
	\bibitem{Costi.00}
	Costi, T.~A., {K}ondo effect in a magnetic field and the magnetoresistivity of
	{K}ondo alloys. \emph{Phys. Rev. Lett.} \textbf{85}, 1504--1507 (2000).
	
	\bibitem{Lien.11}
	Lien, A.-S., Wang, L.~Y., Chu, C.~S. \& Lin, J.-J., Temporal universal
	conductance fluctuations in {RuO$_2$} nanowires due to mobile defects.
	\emph{Phys. Rev. B} \textbf{84}, 155432 (2011).
	
	\bibitem{Affleck.93}
	Affleck, I. \& Ludwig, A.~W., Exact conformal-field-theory results on the
	multichannel {K}ondo effect: Single-fermion {G}reen's function, self-energy,
	and resistivity. \emph{Phys. Rev. B} \textbf{48}, 7297--7321 (1993).
	
	\bibitem{Parcollet.98}
	Parcollet, O. \& Georges, A., Overscreened multichannel {SU(N)} {K}ondo model:
	Large-{N} solution and conformal field theory. \emph{Phys.~Rev.~B}
	\textbf{58}, 3794--3813 (1998).
	
	\bibitem{Cox.93}
	Cox, D. \& Ruckenstein, A., Spin-flavor separation and non-{F}ermi-liquid
	behavior in the multichannel {K}ondo problem: A large-{N} approach.
	\emph{Phys.~Rev.~Lett.} \textbf{71}, 1613--1616 (1993).
	
	\bibitem{Zamani.13}
	Zamani, F., Chowdhury, T., Ribeiro, P., Ingersent, K. \& Kirchner, S., Quantum
	criticality in the two-channel pseudogap {A}nderson model: A test of the
	non-crossing approximation. \emph{Phys. Status Solidi B} \textbf{250},
	547--552 (2013).
	
	\bibitem{Hewson}
	Hewson, A.~C., \emph{The {K}ondo {P}roblem to {H}eavy {F}ermions} (Cambridge
	University Press, Cambridge, 1993).
	
	\bibitem{Kuramoto.16}
	Kuramoto, Y., Composite electronic orders induced by orbital {K}ondo effect.
	\emph{Sci. Bull.} \textbf{61}, 1563 (2016).
	
	\bibitem{Moustakas.96}
	Moustakas, A. \& Fisher, D., {Prospects for non-{F}ermi-liquid behavior of a
		two-level impurity in a metal}. \emph{Phys. Rev. B} \textbf{53}, 4300--4315
	(1996).
	
	\bibitem{Zhu.16}
	Zhu, L.~J., Nie, S.~H., Xiong, P., Schlottmann, P. \& Zhao, J.~H., Orbital
	two-channel {K}ondo effect in epitaxial ferromagnetic {L}1$_0$-{M}n{A}l
	films. \emph{Nat.~Commun.} \textbf{7}, 10817 (2016).
	
	\bibitem{Zhu.17}
	Zhu, L.~J. \& Zhao, J.~H., Anomalous resistivity upturn in epitaxial {{\it
			L}}2$_1$-{C}o$_2${M}n{A}l films. \emph{Sci. Rep.} \textbf{7}, 42931 (2017).
	
	\bibitem{Cichorek.16}
	Cichorek, T., Bochenek, L., Schmidt, M., Czulucki, A., Auffermann, G., Kniep,
	R., Niewa, R., Steglich, F. \& Kirchner, S., Two-Channel {K}ondo Physics due
	to {A}s Vacancies in the Layered Compound {ZrAs$_{1.58}$Se$_{0.39}$}.
	\emph{Phys. Rev. Lett.} \textbf{117}, 106601 (2016).
	
	\bibitem{Gnida.17}
	Gnida, D., Comment on ``Two-Channel {K}ondo Physics due to {A}s Vacancies in
	the Layered Compound {ZrAs$_{1.58}$Se$_{0.39}$}''. \emph{Phys. Rev. Lett.}
	\textbf{118}, 259701 (2017).
	
	\bibitem{Nicklas.12}
	Nicklas, M., Kirchner, S., Borth, R., Gumeniuk, R., Schnelle, W., Rosner, H.,
	Borrmann, H., Leithe-Jasper, A., Grin, Y. \& Steglich, F.,
	Charge-Doping-Driven Evolution of Magnetism and Non-Fermi-Liquid Behavior in
	the Filled Skutterudite
	{${\mathrm{CePt}}_{4}{\mathrm{Ge}}_{12\ensuremath{-}x}{\mathrm{Sb}}_{x}$}.
	\emph{Phys. Rev. Lett.} \textbf{109}, 236405 (2012).
	
	\bibitem{Yamane.18}
	Yamane, Y., Onimaru, T., Wakiya, K., Matsumoto, K.~T., Umeo, K. \& Takabatake,
	T., Single-Site Non-{F}ermi-Liquid Behaviors in a Diluted $4{f}^{2}$ system
	{Y}$_{1-x}${P}r$_x${I}r$_2${Z}n$_{20}$. \emph{Phys. Rev. Lett.} \textbf{121},
	077206 (2018).
	
	\bibitem{Miranda.97}
	Miranda, E., Dobrosavljevi\ifmmode~\acute{c}\else \'{c}\fi{}, V. \& Kotliar,
	G., Disorder-Driven Non-{F}ermi-Liquid Behavior in {K}ondo Alloys.
	\emph{Phys. Rev. Lett.} \textbf{78}, 290--293 (1997).
	
	\bibitem{Morin.59}
	Morin, F.~J., Oxides Which Show a Metal-to-Insulator Transition at the
	{N\'{e}el} Temperature. \emph{Phys. Rev. Lett.} \textbf{3}, 34--36 (1959).
	
	\bibitem{Wall.18}
	Wall, S., Yang, S., Vidas, L., Chollet, M., Glownia, J.~M., Kozina, M.,
	Katayama, T., Henighan, T., Jiang, M., Miller, T.~A., Reis, D.~A., Boatner,
	L.~A., Delaire, O. \& Trigo, M., Ultrafast disordering of vanadium dimers in
	photoexcited {VO$_2$}. \emph{Science} \textbf{362}, 572--576 (2018).
	
	\bibitem{Lai.18}
	Lai, H.-H., Grefe, S.~E., Paschen, S. \& Si, Q., Weyl{\textendash}{Kondo}
	semimetal in heavy-fermion systems. \emph{PNAS} \textbf{115}, 93--97 (2018).
	
	\bibitem{Goh.12}
	Goh, S.~K., Mizukami, Y., Shishido, H., Watanabe, D., Yasumoto, S., Shimozawa,
	M., Yamashita, M., Terashima, T., Yanase, Y., Shibauchi, T., Buzdin, A.~I. \&
	Matsuda, Y., Anomalous Upper Critical Field in {CeCoIn$_5$}/{YbCoIn$_5$}
	Superlattices with a Rashba-Type Heavy Fermion Interface. \emph{Phys. Rev.
		Lett.} \textbf{109}, 157006 (2012).
	
	\bibitem{Simonov.20}
	Simonov, A., De~Baerdemaeker, T., Bostr{\"o}m, H. L.~B., R{\'i}os~G{\'o}mez,
	M.~L., Gray, H.~J., Chernyshov, D., Bosak, A., B{\"u}rgi, H.-B. \& Goodwin,
	A.~L., Hidden diversity of vacancy networks in Prussian blue analogues.
	\emph{Nature} \textbf{578}, 256--260 (2020).
	
	\bibitem{Lai.20}
	Lai, F., Feng, J., Ye, X., Zong, W., He, G., Yang, C., Wang, W., Miao, Y.-E.,
	Pan, B., Yan, W., Liu, T. \& Parkin, I.~P., Oxygen vacancy engineering in
	spinel-structured nanosheet wrapped hollow polyhedra for electrochemical
	nitrogen fixation under ambient conditions. \emph{J. Mater. Chem. A}
	\textbf{8}, 1652--1659 (2020).
	
	\bibitem{Chen.04}
	Chen, R.-S. \& Huang, Y.-S., Field emission from vertically aligned conductive
	{IrO$_2$} nanorods. \emph{Appl. Phys. Lett.} \textbf{84}, 1552 (2004).
	
	\bibitem{Lin.08}
	Lin, Y.~H., Sun, Y.~C., Jian, W.~B., Chang, H.~M., Huang, Y.~S. \& Lin, J.~J.,
	Electrical transport studies of individual {IrO}$_2$ nanorods and their
	nanorod contacts. \emph{Nanotechnology} \textbf{19}, 045711 (2008).
	
	\bibitem{Chen.04b}
	Chen, R., Chang, H., Huang, Y., Tsai, D., Chattopadhyay, S. \& Chen, K., Growth
	and characterization of vertically aligned self-assembled {IrO$_2$} nanotubes
	on oxide substrates. \emph{J. Cryst. Growth} \textbf{271}, 105 -- 112 (2004).
	
	\bibitem{Liu.07}
	Liu, Y.-L., Wu, Z.-Y., Lin, K.-J., Huang, J.-J., Chen, F.-R., Kai, J.-J., Lin,
	Y.-H., Jian, W.-B. \& Lin, J.-J.~L., Growth of single-crystalline {RuO$_2$}
	nanowires with one- and two-nanocontact electrical characterizations.
	\emph{Appl. Phys. Lett.} \textbf{90}, 013105 (2007).
	
	\bibitem{Yeh.17}
	Yeh, S.-S., Chang, W.-Y. \& Lin, J.-J., Probing nanocrystalline grain dynamics
	in nanodevices. \emph{Sci. Adv.} \textbf{3} (2017).
	
	\bibitem{Huang.07}
	Huang, S.~M., Lee, T.~C., Akimoto, H., Kono, K. \& Lin, J.~J., Observation of
	Strong Electron Dephasing in Highly Disordered {Cu$_{93}$Ge$_4$Au$_3$} Thin
	Films. \emph{Phys. Rev. Lett.} \textbf{99}, 046601 (2007).
	
	\bibitem{Momma.11}
	Momma, K. \& Izumi, F., {{\it VESTA3} for three-dimensional visualization of
		crystal, volumetric and morphology data}. \emph{J. Appl. Crystallogr.} \textbf{44}, 1272--1276 (2011).
	
	\bibitem{Costi.94}
	Costi, T.~A., Hewson, A.~C. \& Zlatic, V., Transport coefficients of the
	{A}nderson model via the numerical renormalization group. \emph{J. Phys.:
		Condens. Matter} \textbf{6}, 2519--2558 (1994).
	
	\bibitem{deAlmeida.06}
	de~Almeida, J.~S. \& Ahuja, R., Electronic and optical properties of {RuO$_2$}
	and {IrO$_2$}. \emph{Phys. Rev. B} \textbf{73}, 165102 (2006).
	
\end{thebibliography}

\vskip 21pt

\begin{thebibliography}{10}
	
	\bibitem{Yeh.18s}
	S.-S. Yeh, K.~H. Gao, T.-L. Wu, T.-K. Su, and J.-J. Lin,
	\newblock Phys. Rev. Appl. {\bf 10}, 034004 (2018).
	
	\bibitem{Lin.04s}
	J.~J. Lin et~al.,
	\newblock J. Phys.: Condens. Matter {\bf 16}, 8035 (2004).
	
	\bibitem{Lien.11s}
	A.-S. Lien, L.~Y. Wang, C.~S. Chu, and J.-J. Lin,
	\newblock Phys. Rev. B {\bf 84}, 155432 (2011).
	
	\bibitem{Chiu.09s}
	S.-P. Chiu et~al.,
	\newblock Nanotechnology {\bf 20}, 105203 (2009).
	
	\bibitem{Chiu.13s}
	S.-P. Chiu, J.~G. Lu, and J.-J. Lin,
	\newblock Nanotechnology {\bf 24}, 245203 (2013).
	
	\bibitem{Altshuler.85s}
	B.~L. Altshuler and A.~G. Aronov,
	\newblock {\em Electron-electron interaction in disordered conductors}, book
	section~1,
	\newblock North-Holland Physics Publishing, Amsterdam, The Netherlands, 1985.
	
	\bibitem{Lee.85s}
	P.~A. Lee and T.~V. Ramakrishnan,
	\newblock Rev. Mod. Phys. {\bf 57}, 287 (1985).
	
	\bibitem{Akkermans_s}
	E.~Akkermans and G.~Montambaux,
	\newblock {\em Mesoscopic Physics of electrons and photons},
	\newblock Cambridge University Press, 2006.
	
	\bibitem{Huang.07s}
	S.~M. Huang, T.~C. Lee, H.~Akimoto, K.~Kono, and J.~J. Lin,
	\newblock Phys. Rev. Lett. {\bf 99}, 046601 (2007).
	
	\bibitem{Lin.87s}
	J.~J. Lin and N.~Giordano,
	\newblock Phys. Rev. B {\bf 35}, 545 (1987).
	
	\bibitem{Cichorek.16s}
	T.~Cichorek et~al.,
	\newblock Phys. Rev. Lett. {\bf 117}, 106601 (2016).
	
	\bibitem{Gnida.17s}
	D.~Gnida,
	\newblock Phys. Rev. Lett. {\bf 118}, 259701 (2017).
	
	\bibitem{Cichorek.17s}
	T.~Cichorek et~al.,
	\newblock Phys. Rev. Lett. {\bf 118}, 259702 (2017).
	
	\bibitem{Gnida.18s}
	D.~Gnida,
	\newblock Phys. Rev. B {\bf 97}, 134201 (2018).
	
	\bibitem{Ingold.92s}
	G.-L. Ingold and Y.~V. Nazarov,
	\newblock {\em Charge Tunneling Rates in Ultrasmall Junctions}, book section~1,
	\newblock Plenum Press, New York, 1992.
	
	\bibitem{Furusaki.93s}
	A.~Furusaki and N.~Nagaosa,
	\newblock Phys. Rev. B {\bf 47}, 4631 (1993).
	
	\bibitem{Persson.IrOs}
	K.~Persson,
	\newblock  Materials data on IrO$_2$ (sg:136) by materials project, https://doi.org/10.17188/1201432.
	
	\bibitem{Persson.RuOs}
	K.~Persson,
	\newblock  Materials data on RuO$_2$ (sg:136) by materials project, https://doi.org/10.17188/1307989.
	
	\bibitem{Morgan.10s}
	B.~J. Morgan and G.~W. Watson,
	\newblock J. Phys. Chem. C {\bf 114} (2010).
	
	\bibitem{Lin.15s}
	C.~Lin, D.~Shin, and A.~A. Demkova,
	\newblock J.\,Appl.\,Phys. {\bf 117}, 225703 (2015).
	
	\bibitem{Lechermann.17s}
	F.~Lechermann, W.~Heckel, O.~Kristanovski, and S.~M\"uller,
	\newblock Phys. Rev. B {\bf 95}, 195159 (2017).
	
	\bibitem{Cox.93s}
	D.~Cox and A.~Ruckenstein,
	\newblock Phys.~Rev.~Lett. {\bf 71}, 1613 (1993).
	
	\bibitem{Zhu.19s}
	Z.~H. Zhu et~al.,
	\newblock Phys.\ Rev.\ Lett. {\bf 122}, 017202 (2019).
	
	\bibitem{Berlijn.17s}
	T.~Berlijn et~al.,
	\newblock Phys.\ Rev.\ Lett. {\bf 118}, 077201 (2017).
	
	\bibitem{Parcollet.98s}
	O.~Parcollet and A.~Georges,
	\newblock Phys.~Rev.~B {\bf 58}, 3794 (1998).
	
	\bibitem{Zamani.13s}
	F.~Zamani, T.~Chowdhury, P.~Ribeiro, K.~Ingersent, and S.~Kirchner,
	\newblock physica status solidi (b) {\bf 250}, 547 (2013).
	
	\bibitem{Zamani_thesis_s}
	F.~Zamani,
	\newblock {\em Local quantum criticality in and out of equilibrium},
	\newblock PhD thesis, TU Dresden, 2016.
	
	\bibitem{Costi.94s}
	T.~A. Costi, A.~C. Hewson, and V.~Zlatic,
	\newblock J. Phys.: Condens. Matter {\bf 6}, 2519 (1994).
	
	\bibitem{Choi.06s}
	W.~S. Choi et~al.,
	\newblock Phys. Rev. B {\bf 74}, 205117 (2006).
	
	\bibitem{Kawasaki.18s}
	J.~K. Kawasaki et~al.,
	\newblock Phys. Rev. Lett. {\bf 121}, 176802 (2018).
	
	\bibitem{Affleck.93s}
	I.~Affleck and A.~W. Ludwig,
	\newblock Phys. Rev. B {\bf 48}, 7297 (1993).
	
	\bibitem{Vladar.83s}
	K.~Vlad\'{a}r and A.~Zawadowski,
	\newblock Phys. Rev. B {\bf 28}, 1596 (1983).
	
	\bibitem{Glassford.93s}
	K.~M. Glassford and J.~R. Chelikowsky,
	\newblock Phys. Rev. B {\bf 47}, 1732 (1993).
	
	\bibitem{Shin.97s}
	W.-C. Shin and S.-G. Yoon,
	\newblock J. Electrochem. Soc.  (1997).
	
	\bibitem{Ralph.92s}
	D.~C. Ralph and R.~A. Buhrman,
	\newblock Phys. Rev. Lett. {\bf 69}, 2118 (1992).
	
	\bibitem{Mallet.06s}
	F.~Mallet et~al.,
	\newblock Phys. Rev. Lett. {\bf 97}, 226804 (2006).
	
	\bibitem{Daybell.67s}
	M.~D. Daybell and W.~A. Steyert,
	\newblock Phys. Rev. Lett. {\bf 18}, 398 (1967).
	
	\bibitem{DiTusa.92s}
	J.~F. DiTusa, K.~Lin, M.~Park, M.~S. Isaacson, and J.~M. Parpia,
	\newblock Phys. Rev. Lett. {\bf 68}, 678 (1992).
	
	\bibitem{Chakravarty.00s}
	S.~Chakravarty and C.~Nayak,
	\newblock Int.\ J.\ Mod.\ Phys.\ B {\bf 14}, 1421 (2000).
	
\end{thebibliography}
\end{document}